\documentclass[aps,groupedaddress,nofootinbib,showkeys,showpacs,amsfonts,eqsecnum]{revtex4}

\usepackage{bm}
\usepackage{amsmath}
\newcommand{\tr}{{\rm tr \,}}
\newcommand{\Tr}{{\rm Tr}}
\newcommand{\Det}{{\rm Det\,}}
\newcommand{\Dirac}{{\bf D}}
\newcommand{\Kop}{{\bf K}}

\newcommand{\thru}[1]{\mathrel{\mathop{#1\!\!\!\!/}}}
\newcommand{\thrur}[1]{\mathrel{\mathop{#1\!\!\!/}}}

\newcommand{\D}{D}
\newcommand{\Da}{\hat{D}}
\newcommand{\R}{Q}

\newcommand{\sa}{\mbox{{\textit{a}}}}

\newcommand{\sm}{\mbox{{\textit{m}}}}

\newcommand{\sv}{\mbox{{\textit{v}}}}

\newcommand{\sA}{\mbox{{\textit{B}}}}

\newcommand{\sD}{\mbox{{\textit{D}}}}
\newcommand{\sDa}{\hat{\sD}}
\newcommand{\sF}{\mbox{{\textit{F}}}}

\newcommand{\sK}{\mbox{{\textit{K}}}}

\newcommand{\sN}{\mbox{{\textit{N}}}}

\newcommand{\sR}{\mbox{{\textit{Q}}}}

\newcommand{\fd}{\mbox{\textsf{\textit{d}}}}

\newcommand{\fm}{\mbox{\textsf{\textit{m}}}}

\newcommand{\fv}{\mbox{\textsf{\textit{v}}}}

\newcommand{\fD}{\mbox{\textsf{\textit{D}}}}

\newcommand{\fF}{\mbox{\textsf{\textit{F}}}}
\newcommand{\fJ}{\mbox{\textsf{\textit{J}}}}

\newcommand{\fN}{\mbox{\textsf{\textit{N}}}}

\newcommand{\fR}{\mbox{\textsf{\textit{Q}}}}

\newcommand{\fX}{\mbox{\textsf{\textit{X}}}}

\newcommand{\mlr}{m_{LR}}
\newcommand{\mrl}{m_{RL}}

\newcommand{\m}[1]{{\underline{#1}}}

\begin{document} 

\title{Direct construction of the effective action of chiral gauge
  fermions in the anomalous sector}

\author{L. L. Salcedo}

\affiliation{ Departamento de F{\'\i}sica At\'omica, Molecular y
Nuclear, Universidad de Granada, E-18071 Granada, Spain }

\date{\today}

\begin{abstract}
  The anomaly implies an obstruction to a fully chiral covariant
  calculation of the effective action in the abnormal parity sector of
  chiral theories.  The standard approach then is to reconstruct the
  anomalous effective action from its covariant current.  In this work
  we use a recently introduced formulation which allows to directly
  construct the non trivial chiral invariant part of the effective
  action within a fully covariant formalism. To this end we
  develop an appropriate version of Chan's
  approach to carry out the calculation within the derivative
  expansion.  The result to four derivatives, i.e., to leading order
  in two and four dimensions and next-to-leading order in two
  dimensions, is explicitly worked out. Fairly compact expressions are
  found for these terms.
\end{abstract}



\maketitle


\section{Introduction}
\label{sec:1}

This paper deals with fermions of both chiralities coupled to local
external fields of spin 0 or 1. The physics of such system is
contained in its effective action, formally the logarithm of the
determinant of the Dirac operator \cite{Itzykson:1980bk}. A key
feature of this system is that it enjoys local chiral symmetry at the
classical level. At the quantum level the fermionic measure fails to
be invariant \cite{Fujikawa:1979ay} and this gives rise to the
presence of an anomaly \cite{Adler:1969gk,Bell:1969ts,Bardeen:1969md}
in the abnormal parity sector of the effective action
\cite{AlvarezGaume:1983ig}. The imaginary part of the effective action
contains also other interesting structure, such as many-valuation or
topological terms, and for this reason has been thoroughly studied in
the literature
\cite{Adler:1969er,Wess:1971yu,Goldstone:1981kk,Witten:1983tw,%
  Bardeen:1984pm,D'Hoker:1984ph,Leutwyler:1985em,Alvarez-Gaume:1985dr,%
  Banerjee:1986bu,Ball:1989xg}.  A good account on the
subject can be found in \cite{Ball:1989xg}.

Clearly, the presence of the anomaly in the abnormal parity sector
implies an impediment to a direct construction of the effective action
using a chiral covariant formalism. The anomaly is saturated by the
Wess-Zumino-Witten, so when this term is removed from the effective
action what remains is a chiral invariant functional which can be
written using simple chiral covariant elements, namely, the spin zero
fields, the field strengths and their chiral covariant derivatives.
However, even if this remainder is chiral invariant no direct chiral
covariant construction of it was available in the literature. Instead,
the best route was to compute the chiral covariant version of the
current and then reconstruct the effective action from it
\cite{Salcedo:2000hx,Hernandez:2007ng}.

Recently we have shown \cite{Salcedo:2008tc} that the chiral invariant
remainder can in fact be expressed as the standard effective action of
a local Klein-Gordon operator which moreover is manifestly chiral
invariant.  This opens the possibility to a direct covariant
calculation of the invariant part of the effective action in the
abnormal parity sector, along the same lines available to the real
part. Such calculation is addressed here within the derivative
expansion approach.

In Section \ref{sec:2} we present a background of previous results. In
\ref{subsec:2.A} we briefly recall the concepts involved in
this subject. The main result of \cite{Salcedo:2008tc} is reviewed in
\ref{subsec:2.B}.  Also the calculation in \cite{Salcedo:2008tc} using
the method of Chan \cite{Chan:1986jq} is described in
\ref{subsec:n:2.C}. In \ref{subsec:n:2.D} we introduce some notational
conventions taken from \cite{Salcedo:2000hx}. Such notation is quite
transparent and allows to manipulate the expressions to appear
subsequently suppressing redundant information. Moreover, it permits to
work with formally vector covariant quantities using the original
fields (rather than chirally rotated ones).  \ref{subsec:2.E}
summarizes the available result for the effective action in the
abnormal parity sector at leading order in the covariant derivative
expansion.

In Section \ref{subsec:3.A} we introduce an overcomplete basis of
standard functions to be used subsequently which allow to express in a
simple form the existing results for the effective action or the
current. This is explicitly done in \ref{subsec:3.B} for the leading
order term.  Section \ref{subsec:3.C} is a digression which shows that
the overall structure of the result for the effective action can be
understood in some cases although no simple pattern is found for the
general case.

In Section \ref{sec:4} we apply the findings in \cite{Salcedo:2008tc}
to carry out a direct calculation the effective action at leading
order in the abnormal parity sector. Chan's approach, designed for
bosonic theories, is discussed. This approach can be applied to the
present problem but it implies a redefinition of the covariant
derivative and the algebra quickly becomes quite involved. To bypass
this problem we construct from scratch a completely new derivative
expansion along the same lines of original Chan approach but tailored
for the fermionic case in the abnormal parity sector. The analogous of
Chan's formulas for two derivative terms in two dimensions and four
derivative terms in four dimensions are derived. (Unlike the bosonic
Chan formula, the fermionic one depends on the dimension.) The
calculation is fully worked out to obtain the effective action at
leading order in two and four dimensions and the previous results of
\cite{Salcedo:2000hx} are indeed reproduced.  As a byproduct some
structural properties of the result are found which were not easily
visible in the calculation based on the current.

In Section \ref{sec:5} we discuss the general form of imaginary part
of the effective action to next-to-leading order in the two
dimensional case. In order to compute it we extend our fermionic Chan
formula to four derivatives in two dimensions in \ref{subsec:5.B}. For
comparison, the same calculation using the current method is presented
in \ref{subsec:5.C}. It is verified that both method give the same
result and that this is consistent with the calculation of the same
quantity in \cite{Hernandez:2007ng}, where the world-line approach is
used.

Section \ref{sec:6} summarizes our conclusions. The relation between
the basis of functions introduced in \ref{subsec:3.A} and the usual
momentum integrals is established in the Appendix.

\section{Background of  previous results}
\label{sec:2}

\subsection{The Dirac operator and the effective action}
\label{subsec:2.A}

The Dirac operators we consider describe Dirac fermions coupled to
general spin 0 and 1 fields without derivative couplings. These are of
the form
\begin{equation}
\Dirac= \thru{\D}_R P_R \, +\thru{\D}_L P_L + \mlr P_R+\mrl P_L \,,
\label{eq:n:1}
\end{equation}
with $D_\mu^{R,L}=\partial_\mu+v_\mu^{R,L}$,
$P_{R,L}=\frac{1}{2}(1\pm\gamma_5)$.  $d$ is the (even) dimension of
the Euclidean space-time and the background fields $v^{R,L}_\mu(x)$
and $\mlr(x)$, $\mrl(x)$ are matrices in some arbitrary internal
space. Unitarity requires $v^{L,R}_\mu$ to be antihermitian and
$\mrl^\dagger=\mlr$. In addition, we assume that the matrices $\mlr$,
$\mrl$ are not singular at any point. The space-time is flat and the
temperature zero.

The fermionic effective action $W$ is introduced through standard
functional integration of the fermionic fields
\begin{equation}
e^{-W}=\int {\cal D}\bar\psi {\cal D}\psi \,e^{-\int d^dx \,\bar\psi \Dirac \psi} 
= \Det \Dirac
\end{equation}
so formally
\begin{equation}
W= -\Tr\,\log\Dirac
\end{equation}
modulo ultraviolet (UV) ambiguities. ($\Tr$ represents the functional
trace.) 

$W$ can be split into normal and abnormal parity components,
$W=W^++W^-$. $W^+$ is the component without Levi-Civita pseudo-tensor,
is real (in Euclidean space) and even under the exchange
$R\leftrightarrow L$. $W^-$ contains the Levi-Civita pseudo-tensor, is
imaginary and odd under under the exchange of chiral labels
$R\leftrightarrow L$. In this work we concentrate on the abnormal
parity component, $W^-$. In general we will follow the notation and
conventions of \cite{Salcedo:2000hp,Salcedo:2000hx,Salcedo:2008tc} to
which we refer for further details. In particular, the Dirac gammas
are hermitian and
\begin{equation}
\gamma_\mu\gamma_\nu=\delta_{\mu\nu}+\sigma_{\mu\nu}
,
\qquad
\gamma_5=i^{d/2}\gamma_0\cdots\gamma_{d-1}
.
\end{equation}

The fermionic action $\int d^dx \,\bar\psi \Dirac \psi$ is invariant
under local chiral transformations
\begin{eqnarray}
\psi^\Omega &=& 
(\Omega_R^{-1}P_R+\Omega_L^{-1}P_L)\psi
,
\qquad
\bar\psi^\Omega=\bar\psi(\Omega_L P_R+\Omega_R P_L)
\nonumber \\
\Dirac^\Omega &=&
\thru{\D}{}^\Omega_R P_R \, +\thru{\D}{}^\Omega_L P_L 
+ \mlr^\Omega P_R+\mrl^\Omega P_L \,
\end{eqnarray}
with
\begin{equation}
(v_\mu^{R,L})^\Omega= \Omega_{R,L}^{-1}v_\mu^{R,L} \Omega_{R,L} + 
\Omega_{R,L}^{-1}[\partial_\mu,\Omega_{R,L}]
\,,\qquad
\mlr^\Omega= \Omega_L^{-1}\mlr \Omega_R
\,,\qquad
\mrl^\Omega= \Omega_R^{-1}\mrl \Omega_L
\,,
\label{eq:n:2.6}
\end{equation}
and so, $(D_\mu^{R,L})^\Omega= \Omega_{R,L}^{-1}
D_\mu^{R,L}\Omega_{R,L}$. Chiral covariant derivatives and field
strengths are defined correspondingly:
\begin{eqnarray}
(\Da_\mu m)_{RL} &=& 
D^R_\mu\mrl-\mrl D^L_\mu \,,
\quad
(\Da_\mu m)_{LR} = D^L_\mu\mlr-\mlr D^R_\mu \,,
\quad
F^{R,L}_{\mu\nu} =
[D^{R,L}_\mu,D^{R,L}_\nu]
.
\label{eq:n:2.7}
\end{eqnarray}

In general the effective action cannot be determined in closed form
and so several expansions are used to address its systematic
calculation. Of special interest to us will be the derivative
expansion. In this approach the terms are classified by the number of
covariant derivatives they carry. Features of this expansion are that
different orders do not mix under chiral rotations and also that UV
ambiguities affect only terms with $d$ derivatives or less.  The
derivative expansion starts at order $d$ ($d$ derivatives) in the
abnormal parity sector.

\subsection{The invariant factor of the chiral determinant}
\label{subsec:2.B}

In this section we summarize the findings in \cite{Salcedo:2008tc}.

As is well known the symmetry under chiral gauge transformations is
broken by an anomaly which can be eliminated in $W^+$ but not in
$W^-$. The chiral anomaly is saturated by the gauged
Wess-Zumino-Witten (WZW) action, so, in the abnormal parity sector,
the effective action can be written as
\begin{equation}
W^- = \Gamma_{\rm gWZW}+W^-_c \,,
\end{equation}
where $\Gamma_{\rm gWZW}$ is the gauged WZW action
\cite{Witten:1983tw} and $W^-_c$ is the chiral invariant remainder. As
it stands there is an ambiguity in the separation between anomalous
and non anomalous pieces. This ambiguity is resolved in
\cite{Salcedo:2000hx,Salcedo:2008tc} showing that there is a natural
choice of $\Gamma_{\rm gWZW}$. For instance in two dimensions:
\begin{eqnarray}
\Gamma_{\text{gWZW}} &=&
-\frac{i}{24\pi}\int\epsilon_{\mu\nu\alpha}
\tr\big(
\mlr^{-1}\partial_\mu\mlr \,
 \mlr^{-1}\partial_\nu\mlr \,
\mlr^{-1}\partial_\alpha\mlr 
-
\mrl^{-1}\partial_\mu\mrl \,
 \mrl^{-1}\partial_\nu\mrl \,
\mrl^{-1}\partial_\alpha\mrl
\big)
\, d^3x
\nonumber \\
&&
+\frac{i}{8\pi}\int\epsilon_{\mu\nu}
\tr\big(
\partial_\mu\mrl \, \mrl^{-1} v^R_\nu
-\partial_\mu\mlr \, \mlr^{-1} v^L_\nu
-\mlr^{-1}\partial_\mu\mlr  \, v^R_\nu
+\mrl^{-1}\partial_\mu\mrl  \, v^L_\nu
\nonumber \\
&& \qquad\quad
-\mrl \,v^L_\mu \mrl^{-1} v^R_\nu
+\mlr \,v^R_\mu \mlr^{-1} v^L_\nu
\big)
\, d^2x
.
\label{eq:n:2.9a}
\end{eqnarray}

The functional $W^-_c$ is chiral invariant and remarkably it can be
expressed as the $\Tr\,\log{}$ of a Klein-Gordon like operator. Indeed,
as shown in \cite{Salcedo:2008tc}
\begin{equation}
W_c = 
-\frac{1}{2}\Tr\,\log
\Kop
,
\label{eq:n:2.9}
\end{equation}
where
\begin{equation}
\Kop = K_L P_R+ K_R P_L
\end{equation}
and
\begin{equation}
K_R = \mrl\mlr \, - \thru{\D}_R  \mlr^{-1} \thru{\D}_L \mlr
,
\qquad
K_L=\mlr\mrl - \thru{\D}_L \mrl^{-1} \thru{\D}_R \mrl
.
\label{eq:n:2.12}
\end{equation}

In (\ref{eq:n:2.9}) $W_c$ refers to $W^+$, which is
chiral invariant, plus $W^-_c$. It implies
\begin{eqnarray}
W^+ &=& -\frac{1}{4}\Tr\log K_R-\frac{1}{4}\Tr\log K_L
,
\\
W^-_c &=& \frac{1}{4}\Tr\left(\gamma_5\log K_R\right)
-\frac{1}{4}\Tr\left(\gamma_5\log K_L\right)
.
\label{eq:n:2.13}
\end{eqnarray}

Let us explain why (\ref{eq:n:2.9}) is of interest. As noted,
the real part of the effective action $W^+$ is chiral invariant and
can be computed by means of
\begin{equation}
W^+ = -\frac{1}{2}\Tr\log(\Dirac^\dagger\Dirac)
\,.
\end{equation}
$\Dirac^\dagger\Dirac$ has all the good properties. It is a positive
definite Klein-Gordon operator and so there is a variety of methods
available in the bosonic case to work out the computation, in
particular the heat kernel approach. Chiral invariance helps also
since everything will depend on chiral covariant blocks, namely,
$\mlr$, $\mrl$, $F^{R,L}_{\mu\nu}$ and their covariant derivatives.
For the imaginary part, one has instead
\begin{equation}
W^- = -\frac{1}{2}\Tr\log(\Dirac^\dagger{}^{-1}\Dirac)
\,.
\end{equation}
The operator $\Dirac^\dagger{}^{-1}\Dirac$ does not enjoy these nice
features. In fact it is non local and non chiral covariant. An obvious
approach is to use $\Dirac^2$ (assuming a suitable analytic
continuation to turn $\Dirac$ into an Hermitian matrix), so that
\begin{equation}
W = -\frac{1}{2}\Tr\log(\Dirac^2)
\,.
\end{equation}
$\Dirac^2$ has the virtue of being of the Klein-Gordon type.
Unfortunately $\Dirac^2$ is far from transforming nicely under chiral
rotations. As a consequence only vector gauge invariance is preserved
and so calculations along this route are difficult \cite{Dhar:1985gh}.
A direct computation of $\Tr\log(\Dirac)$ (rather than
$\Tr\log(\Dirac^2)$) was undertook in \cite{Salcedo:1996qy} using the
$\zeta$-function approach \cite{Hawking:1977ja,Elizalde:1994bk}. Once
again only vector gauge invariance is preserved in this approach.

At the root of these problems is the presence of the chiral anomaly in
$W^-$.  The anomaly is an obstruction to a chiral invariant
treatment. Any approach that tries to compute $W^-$ (or $W$) without
previous extraction of the anomaly cannot enjoy chiral invariance.

A chiral covariant approach is based on the current. Unlike the
effective action, the current (the variation of $W$ with respect to
the gauge fields) admits a chiral covariant version. This is the so
called covariant current. Such current is not directly consistent
(i.e., it is not a true variation) but it can be turned into a
consistent current by adding the appropriate counter-term
\cite{Bardeen:1984pm}. The point is that the covariant current is
amenable to direct chiral covariant computation. Then $W_c^-$,
constructed with covariant blocks, can be adjusted in order to
reproduce the current. This is the approach introduced in
\cite{Salcedo:2000hx} and applied there to compute $W^-_c$ at leading
order of the derivative expansion in two and four dimensions. The same
method has been applied in \cite{Hernandez:2007ng} to compute the
next-to-leading order in two dimensions.

The abovementioned obstruction induced by the anomaly is a consequence
of the ambiguities introduced by the ultraviolet divergences, as is
the mismatch between the covariant and consistent currents. Therefore
there should be no obstruction in ultraviolet finite terms. This
includes all terms beyond the lowest order one in the derivative
expansion of $W^-$.  In this view, the current method of
\cite{Salcedo:2000hx} is covariant but is not a direct computation of
the effective action.

The relation (\ref{eq:n:2.9}), and in particular (\ref{eq:n:2.13}),
provides us precisely with such a direct approach to $W^-_c$ (to all
orders actually). $\Kop$ is a manifestly chiral covariant operator of
the Klein-Gordon type and this opens the possibility to address the
computation of the imaginary part of the effective action $W^-_c$
using the same efficient techniques available for the real part. 

It should be noted that the calculation of $W_c$ through $\Tr\log\Kop$
is still subject to UV ambiguities. These are chiral covariant
polynomial counter-terms constructed with $D^{R,L}_\mu$,
$\mlr^{-1}(\Da_\mu m)_{LR}$, and $\mrl^{-1}(\Da_\mu m)_{RL}$ in
$W^-_c$ and also $\mrl\mlr$, $\mlr\mrl$ in $W^+$.  Some of these
polynomial are spurious contributions to $W_c$ and should be removed
\cite{Salcedo:2008tc}.

\subsection{Chan-like calculation}
\label{subsec:n:2.C}

The operators $K_{R,L}$ can be brought to a standard Klein-Gordon form
\begin{eqnarray}
K_R
&=&
{\tilde M}_R - ({\tilde D}^R_\mu)^2 
,
\qquad
K_L
=
{\tilde M}_L - ({\tilde D}^L_\mu)^2 
\end{eqnarray}
with
\begin{eqnarray}
{\tilde M}_R &=& \mrl\mlr
-\frac{1}{2}\sigma_{\mu\nu} F^R_{\mu\nu}
+ (B^R_\mu)^2 -[D^R_\mu, B^R_\mu] \,,
\nonumber \\
{\tilde D}^R_\mu &=& D^R_\mu+B^R_\mu\,,
\nonumber \\
B^R_\mu &=& 
\frac{1}{2}\gamma_\mu\gamma_\nu \,\R^R_\nu
\,.
\label{eq:4.10}
\end{eqnarray}
(and similarly for $K_L$)
where
\begin{equation}
\R^R_\mu = \mlr^{-1}(\Da_\mu m)_{LR}
,
\qquad
\R^L_\mu = \mrl^{-1}(\Da_\mu m)_{RL}
.
\end{equation}

The standard form ``$M-D_\mu^2$'' allows to apply Chan's method
\cite{Chan:1986jq} straightforwardly. This calculation has been carried out in
\cite{Salcedo:2008tc} for $W^-_c$ to two derivatives in two
dimensions and yields
\begin{eqnarray}
W^-_{c,d=2,\text{LO}} &=&
-\frac{i}{2}\int \frac{d^2x d^2p}{(2\pi)^2}\epsilon_{\mu\nu}\tr\!\Big(
  N_R^2 \mrl (\Da_\mu m)_{LR}  N_R \mrl  (\Da_\nu m)_{LR}
\nonumber \\
&&
- N_L^2 \mlr (\Da_\mu m)_{RL}  N_L \mlr  (\Da_\nu m)_{RL}
\Big)
\label{eq:n:2.21}
\end{eqnarray}
where
\begin{equation}
N_R= (p^2+\mrl\mlr)^{-1}
,
\qquad
N_L= (p^2+\mlr\mrl)^{-1}
.
\end{equation}
The label LO refers to leading order in the derivative expansion. As
shown in \cite{Salcedo:2008tc} (\ref{eq:n:2.21}) leads to the result
previously established in \cite{Salcedo:2000hx}.

\subsection{Notational conventions}
\label{subsec:n:2.D}

As we have just noted, it is possible to apply (\ref{eq:n:2.13}) to
compute $W^-_c$ using only chiral covariant quantities. It is also
clear that as one considers more complicated cases (e.g. four
derivatives) the formulas will become more involved. This leads us to
use a number of notational conventions in order to simplify the
expressions. Such conventions where already introduced in
\cite{Salcedo:2000hp,Salcedo:2000hx}.

For convenience, in what follows we will use Lorentz indices to
indicate covariant derivatives, e.g.,
\begin{eqnarray}
m^{LR}_{\mu\nu}= \Da_\mu\Da_\nu m_{LR}
\,,\quad
F^R_{\alpha\mu\nu}=\Da_\alpha F^R_{\mu\nu} \,.
\end{eqnarray}
(Each new derivative adds an index to left.)

Under chiral transformations, the various quantities transform as $LR$
(e.g. $\mlr$), $RL$, $RR$ (e.g., $D_\mu^R$ or $F^R_{\mu\nu}$) or $LL$.
However, due to chiral symmetry, it is clear that the explicit writing
of these labels is largely redundant.  In fact, as can be seen in
previous formulas, in functionals like the effective action, a field
with a label $R$ at the right (i.e., of type $LR$ or $RR$ under chiral
transformations) must be followed by a field with a label $R$ at the
left (i.e., of type $RL$ or $RR$).  Also, expressions inside the trace
should start and end with the same label, $R$ or $L$. In addition,
$W^-$ is odd under the exchange of chiral labels $R\leftrightarrow L$.
This implies that we can work in the abnormal parity sector without
explicitly using chiral labels: a traced expression without labels
will mean half the expression starting with $R$ minus half the
expression starting with $L$. E.g.
\begin{eqnarray}
\tr(\sF\!{}_{\mu\nu}\,\sm_\mu\,\sm_\nu) &=&
\frac{1}{2}\tr(F_{\mu\nu}^R  \, m^{RL}_\mu  m^{LR}_\nu) -
\frac{1}{2}\tr(F_{\mu\nu}^L  \, m^{LR}_\mu  m^{RL}_\nu)\,.
\label{eq:2.4}
\end{eqnarray}
($\tr$ denotes the trace over internal degrees of freedom.) There is
only one subtlety to be noted: within this notation, quantities that
flip the chiral label, such $\sm$ or $\sm_\mu$, are {\em odd} under
the cyclic property of the trace:
\begin{eqnarray}
\tr(\sF\!{}_{\mu\nu}\,\sm_\mu\,\sm_\nu) &=&
\frac{1}{2}\tr(m^{LR}_\nu F_{\mu\nu}^R \, m^{RL}_\mu  ) -
\frac{1}{2}\tr(m^{RL}_\nu F_{\mu\nu}^L  \,m^{LR}_\mu  )
= -\tr(\sm_\nu\sF\!{}_{\mu\nu}\,\sm_\mu)
\,.
\label{eq:2.5}
\end{eqnarray}
Note also that inside the trace the total number of $\sm$'s (with or
without derivatives) should always be even.  This is because there
should be as many $LR$ fields as $RL$ ones (since expressions in the
trace start and end with the same label).

For instance, without any loss of information (\ref{eq:n:2.6}) and
(\ref{eq:n:2.7}) can be written as
\begin{eqnarray}
v_\mu^\Omega &=& \Omega^{-1}v_\mu \Omega + 
\Omega^{-1}[\partial_\mu,\Omega]
\,,\qquad
\sm^\Omega= \Omega^{-1} \sm \Omega
\,,\qquad
D_\mu^\Omega= \Omega^{-1} D_\mu\Omega
,
\nonumber \\
\Da_\mu m &=& 
D_\mu \sm - \sm D_\mu 
= [D_\mu, \sm ]
=\sm_\mu
\,,
\qquad
F_{\mu\nu} =
[D_\mu,D_\nu]
.
\label{eq:n:2,26}
\end{eqnarray}
Also (\ref{eq:n:2.12}) and (\ref{eq:n:2.13}) can be written as
\begin{equation}
\sK = \sm^2 - \thru{\D} \sm^{-1}\! \thru{\D} \sm
,
\qquad
W^-_c = \frac{1}{2}\Tr\left(\gamma_5\log \sK\right)
,
\label{eq:n:2.27}
\end{equation}

Likewise (\ref{eq:n:2.9a}) becomes\footnote{The gauged WZW term can
  also be written using this notation since it relies only on
  invariance under global chiral rotations.}
\begin{equation}
\Gamma_{\rm gWZW} =
   -\frac{i}{12\pi}\int\tr (\sm^{-1} \fd \sm)^3 
+ 
   \frac{i}{4\pi}\int\tr
\left(
   \fd \sm \,\sm^{-1}\fv 
- \sm^{-1} \fd \sm \,\fv - \sm\fv\sm^{-1}\fv
 \right)
\,.
\end{equation}
Here $\fd \sm= \partial_\mu \sm \fd x_\mu$, $\fv= \sv_\mu \fd
x_\mu$. In this formula we use differential forms. Differential forms
are introduced in the formulas through the identity
\begin{equation}
\epsilon_{\mu_1\mu_2\cdots\mu_d}d^d x = \fd x_{\mu_1}\fd
x_{\mu_2}\cdots \fd x_{\mu_d}
.
\end{equation}
We will also use the differential forms
\begin{equation}
\fm{}^\prime=\sm_\mu \fd x_\mu \,, \quad
\fF=\frac{1}{2}\sF\!{}_{\mu\nu} \fd x_\mu \fd x_\nu \,,
\end{equation}

The notation 
\begin{equation}
\fX{}^\prime:= \fd x_\mu\sDa_\mu\fX
\end{equation}
will be used throughout. Here $\fX$ represents any matrix-valued
$n$-form. Two useful results are $\fX^{\prime\prime}=[\fF,\fX]$ and
$\fF^\prime=0$.

Finally, (\ref{eq:n:2.21}) becomes
\begin{eqnarray}
W^-_{c,d=2,\text{LO}} &=&
-i\int \frac{d^2p}{(2\pi)^2}\tr\!\left(
  N^2 \sm \fm^\prime   N \sm  \fm^\prime
\right)
\label{eq:n:2.21a}
\end{eqnarray}
where
\begin{equation}
N= (p^2+\sm^2)^{-1}
.
\end{equation}

A further notational convention is that of labeled operators
\cite{Feynman:1951gn,Garcia-Recio:2000gt,Salcedo:2000hp,%
Salcedo:2000hx,Salcedo:2004yh}:
we will use labels $1,2,3,\ldots$ in the quantity $\sm$ (with no
derivatives) to indicate its position in an expression. For instance
\begin{equation}
\sm^2\fF \sm\fF\sm^{-1}\fm^\prime\sm^3
= \frac{\sm_1^2\sm_2\sm_4^3}{\sm_3} \fF^2 \fm^\prime \,.
\end{equation}
This is useful because the labeled $\sm$'s can be treated as
c-numbers, and so compact expressions like $f(\sm_1,\sm_2)\fF$ become
meaningful ($f(x,y)$ being an ordinary function). As discussed in
\cite{Salcedo:2000hx}, an expression like $f(\sm_1,\sm_2)\fF$ is well
defined provided $f(x,y)$ is regular in the coincidence limit
$x^2-y^2\to 0 $. (On the other hand, the use of labeled operators
with functions violating the regularity condition in the coincidence
limit may yield nonsensical expressions \cite{Salcedo:2000hx}.)

Using this notation (\ref{eq:n:2.21a}) becomes
\begin{eqnarray}
W^-_{c,d=2,\text{LO}} &=&
-i\int \frac{d^2p}{(2\pi)^2}\tr\!\left[
  \frac{\sm_1 \sm_2}{(p^2+\sm_1^2)^2(p^2+\sm_2^2)}  \fm^\prime \fm^\prime
\right]
.
\end{eqnarray}
The point of labeling the operators is that the momentum integration
is now straightforward and yields
\begin{eqnarray}
W^-_{c,d=2,\text{LO}} &=&
\frac{i}{4\pi}\int \frac{\sm_1 \sm_2 }{m_1^2-m^2_2}\left[
\frac{1}{2}\left(\frac{1}{m_1^2}+\frac{1}{m_2^2}\right)
-\frac{\log(m^2_1/m^2_2)}{m^2_1-m^2_2}
\right]
\fm^\prime \fm^\prime
.
\label{eq:n:2.36}
\end{eqnarray}
This coincides with (3.41) of \cite{Salcedo:2008tc} where the
expression is given using the eigen-basis method instead of labeled
operators.

A noteworthy feature of our conventions is that within this notation
chiral invariance becomes formally identical to vector invariance (cf.
(\ref{eq:n:2,26})). This is different from the standard approach of
chirally rotating the background fields $v^{R,L}_\mu$, $\mlr$, $\mrl$
so that the rotated $\mlr$, $\mrl$ become equal. With that choice the
remaining freedom is vector gauge invariance. By construction, in that
approach any vector invariant calculation becomes chiral invariant
upon undoing the chiral rotation. However, the rotated expressions
depend on three fields, $V_\mu=(v_\mu^R+v_\mu^L)/2$,
$A_\mu=(v_\mu^R-v_\mu^L)/2$, and $S=\mlr=\mrl$, instead of only two,
$\sv_\mu$ and $\sm$. We achieve formal vector gauge invariance using
the original fields, without any change of variables to rotated
variables.

We have introduced three main conventions here, namely, (i) removing
the redundant chiral labels, (ii) antisymmetrization with respect to
$R,L$ labels in traced quantities and (iii) labeling $m$ with respect
to fixed operators.  Perhaps some readers may find this notation
obscure. On the contrary, we think that our conventions highlight the
underlying structure of the expressions, are quite natural and well
suited to the present context and no ambiguity is introduced.  In any
case, it is a fact that the LO terms of $W^-$ at $d=2,4$ for a generic
theories were not obtained until this notation was used in
\cite{Salcedo:2000hx}.

\subsection{The effective action in the abnormal parity sector}
\label{subsec:2.E}

At leading order, the remainder $W^-_c$ vanishes identically when the
scalar and pseudo-scalar fields satisfy a generalized chiral circle
constraint, namely, when $\mrl\mlr$ is a c-number. However, in
general, $W^-_c$ is a non trivial functional.  $W_c^-$ has been
computed in \cite{Salcedo:2000hx} for $d=2,4$ to leading order (LO),
that is, to $d$ covariant derivatives, and next-to-leading order (NLO)
in \cite{Hernandez:2007ng} for $d=2$.  The LO results take the form
\begin{eqnarray}
W^-_{c,{\rm LO}} &=& \left\langle N_{12}\fm^\prime{}^2 \right\rangle 
  \qquad\qquad\qquad\quad  (d=2) \nonumber \\
W^-_{c,{\rm LO}} &=&  \left\langle N_{1234}\fm{}^\prime{}^4 + 
N_{123}\fm{}^\prime{}^2\fF \right\rangle \quad (d=4) \,.
\label{eq:2.9}
\end{eqnarray}

In these formulae the symbol $\langle~\rangle$ is a shorthand for
\begin{equation}
\langle \fX \rangle := 
\frac{i^{d/2} (d/2)!}{(2\pi)^{d/2}d!}\int\tr(\fX)\,
\label{eq:n2}
\end{equation}
($\fX$ being a $d$-form). 

On the other hand $N_{12}=N(\sm_1,\sm_2)$,
$N_{123}=N(\sm_1,\sm_2,\sm_3)$, $N_{1234}=N(\sm_1,\sm_2,\sm_3,\sm_4)$
are three known functions of the labeled $\sm$'s
\cite{Salcedo:2000hx}. For instance,
\begin{eqnarray}
N_{12} &=&
-\frac{\sm_1\sm_2}{\sm_1^2-\sm_2^2}\left(\frac{\log(\sm_1^2/\sm_2^2)}
{\sm_1^2-\sm_2^2}-\frac{1}{2}\left(
\frac{1}{\sm_1^2}+\frac{1}{\sm_2^2}\right)\right)
\,.
\label{eq:n13a}
\end{eqnarray}
(This is just the result quoted in (\ref{eq:n:2.36}).)

Terms of the form $\left\langle f(\sm)\fF \right\rangle$ for $d=2$ or
$\left\langle f(\sm_1,\sm_2)\fF^2 \right\rangle$ for $d=4$ do not
appear in (\ref{eq:2.9}) because they can be removed by using
integration by parts. On the other hand, there is an ambiguity in the
functions $N_{123}$ and $N_{1234}$ of $d=4$ due to the identity
\begin{equation}
0=  \left\langle \left(H_{123}\fm{}^\prime{}^3\right)^\prime\right\rangle
\label{eq:n16}
\end{equation}
and $\fm{}^\prime{}^\prime=[\fF,\sm]$, for any function $H_{123}$.

Cyclic symmetry of the trace allows to impose the conditions
\begin{equation}
N_{12}= N_{2\m{1}}\,, \quad N_{1234}= N_{234\m{1}}
\,,\quad
H_{123}=  -H_{23\m{1}} 
\,,
\label{eq:n3}
\end{equation}
where we use the short-hand $N_{2\m{1}}=N(\sm_2,-\sm_1)$, etc.
Further, unitarity guarantees the following {\em mirror symmetry}
\cite{Salcedo:2000hx}:
\begin{equation}
N_{12} = -N_{21}\,,\quad 
N_{123} = -N_{321}\,,\quad 
N_{1234} = -N_{4321}\,,\quad 
H_{123}= -H_{321}\,.
\label{eq:2.13}
\end{equation}

This ends our summary of previous results.

\section{The chiral remainder at  LO}
\label{sec:3}

\subsection{Set of standard functions}
\label{subsec:3.A}

In \cite{Salcedo:2000hx} the $d=4$ functions $N_{123}$ and $N_{1234}$
were obtained by means of a rather tortuous procedure. If these
functions were unique they would certainly satisfy the condition of
regularity in the coincidence limit ($\sm_i^2\to\sm_j^2$) noted in
Section \ref{subsec:n:2.D} for the use of labeled operators. However,
the existence of the ambiguity (\ref{eq:n16}) allowed spurious
solutions violating that condition.  As a consequence some ingenuity
was required to properly fix the ambiguity in that approach. As we
will see in Section \ref{sec:4}, the method based on (\ref{eq:n:2.13})
directly yields acceptable results.  Also the explicit results in
\cite{Salcedo:2000hx} were complicated and lacked any systematics.

Here we present simple expressions for $N_{12}$, $N_{123}$ and
$N_{1234}$ in which the regularity condition is manifestly checked.
This is achieved by making use of the set of ``standard'' functions
\begin{equation}
  I^\alpha_{r_1,\ldots,r_n} := \oint \frac{dz}{2\pi i} \,z^\alpha \log(z) 
\prod_{j=1}^n\frac{1}{(z-\sm_j^2)^{r_j}} \,,
\quad r_j\in\mathbb{Z}\,,\quad \alpha\in\mathbb{C} \,,
\end{equation}
where the integration is along a positive closed simple contour
enclosing the poles at $\sm_j^2$ but excluding $z=0$. In applications
the $\sm_j^2$ are positive and we choose $\log(z)$ real on the real
positive axis with the branch cut taken along the negative real axis.
Besides $z^\alpha=\exp(\alpha\log(z))$. On their Riemann surfaces the
functions $ I^\alpha_{r_1,\ldots,r_n}$ (as functions of the $m_j^2$)
are regular everywhere, except at $\sm_j=0$ where they present
branching points or poles. In particular they are manifestly regular
in the coincidence limits $\sm^2_i\to\sm^2_j$. They are easy to
compute by residues. In addition, different values of $n$ are related
by recurrence relations and increasing values of $r_j$ are can be
obtained through derivatives with respect to $\sm_j^2$
\cite{Salcedo:2004yh}. For instance
\begin{eqnarray}
I^0_1 &=& \log(\sm_1^2)
,
\nonumber \\
I^0_{2,1} &=& \frac{1}{m_1^2}\frac{1}{m_1^2-m_2^2}
-\frac{\log(\sm_1^2)-\log(\sm_2^2)}{(m_1^2-m_2^2)^2}
.
\end{eqnarray}

These functions are introduced in the calculation in a natural manner
due to the following relation, which holds provided $s+d/2-1$ is a non
negative integer (see the Appendix),
\begin{equation}
\int \frac{d^dp}{(2\pi)^d}(p^2)^s\prod_{j=1}^n\frac{1}{(p^2+\sm_j^2)^{r_j}}
=
\frac{(-1)^{s+d/2-1+\sum_j r_j}}{(4\pi)^{d/2}\Gamma(d/2)}
I^{s+d/2-1}_{r_1,\ldots,r_n}
, 
\qquad
s+d/2=1,2,3,\ldots
\label{eq:3.2}
\end{equation}
This relation holds modulo possible UV divergent contributions on the
left-hand side; the right-hand side is always finite. A further
convenient feature of the standard functions $I^\alpha_{r_1,\ldots,r_n}$ is
that they do not depend on the space-time dimension $d$.

As an application, we write in this basis the functions introduced in
Eqs. (34) and (36) of \cite{Salcedo:2000hx}, corresponding to the
covariant current at LO in two and four dimensions:
\begin{eqnarray}
A_{12} &=& -2 \sm_1 I^1_{2,1}+2 \sm_2 I^1_{1,2} \,,
\nonumber \\
A_{123} &=& 6 \sm_1 I^2_{2,1,1}+6 \sm_2 I^2_{1,2,1}-6 \sm_3 I^2_{1,1,2} \,,
\nonumber \\
A_{1234} &=& -6 \sm_1 I^2_{2,1,1,1}+6 \sm_2 I^2_{1,2,1,1}-6 
\sm_3 I^2_{1,1,2,1}+6 \sm_4 I^2_{1,1,1,2} \,.
\label{eq:3.3a}
\end{eqnarray}

\subsection{LO terms using standard functions}
\label{subsec:3.B}

In order to use the basis $I^\alpha_{r_1,\ldots,r_n}$ to express the functions
$N_{12}$, $N_{123}$ and $N_{1234}$, we first decompose the latter in
components with well-defined parity under $\sm_j\to -\sm_j$, for each
$j$. For instance,
\begin{equation}
N_{12}=
N_{12}^{++}+
\sm_1 N_{12}^{-+}+
\sm_2 N_{12}^{+-}+
\sm_1 \sm_2 N_{12}^{--},
\label{eq:3.3}
\end{equation}
with components $N_{12}^{\pm\pm}$ depending on $\sm_1^2$ and $\sm_2^2$.
Then the $d=2$ result in (\ref{eq:n13a}) can be written as
\begin{equation}
N_{12}^{--}=\frac{1}{2}(I^0_{2,1}-I^0_{1,2})\,, \quad
N_{12}^{++}=N_{12}^{-+}=N_{12}^{+-}=0 \,.
\label{eq:3.4}
\end{equation}
Note that the functions $I^\alpha_{r_1,\ldots,r_n}$ are not linearly
independent and a single function, such as $N_{12}^{--}$, can be
written in different ways using them.

In the case $d=4$, the functions $N_{123}$ and $N_{1234}$ are the
solutions of the Eqs. (90) of \cite{Salcedo:2000hx}. As noted, these
functions are not unique due the ambiguity introduced by
(\ref{eq:n16}). One of the solutions is presented in
\cite{Salcedo:2000hx}, however, we have not tried to express this
particular solution in terms of the basis $I^\alpha_{r_1,\ldots,r_n}$.
Instead, we have directly used the defining equations and rewritten
them in terms of the components $N_{123}^{\pm\pm\pm}$ and
$N_{1234}^{\pm\pm\pm\pm}$.

Because the functions $N_{123}$ and $N_{1234}$ are even under
$\sm\to-\sm$, all the odd components (e.g. $N_{123}^{-++}$ or
$N_{123}^{---}$) vanish.  The ambiguity introduced by (\ref{eq:n16})
can be used to set $N_{123}^{-+-}$ to zero and in this case
$N_{123}^{+++}$ also turns out to be zero. The remaining components of
$N_{123}$ are related by mirror symmetry (\ref{eq:2.13}), namely,
$N_{123}^{+--}=-N_{321}^{--+}$.  For the latter we find the following
valid choice (the ambiguity was not completely fixed by our
previous choice $N_{123}^{-+-}=0$)
\begin{equation}
  N_{123}^{--+}= 6I^1_{1,2,1}
  +\frac{1}{2}\sm_3^2 \left( I^0_{2,1,1}-3 I^0_{1,2,1} +5 I^0_{1,1,2} \right)
.
\label{eq:n:3.7}
\end{equation}

Once the ambiguity has been settled for $N_{123}$, the function
$N_{1234}$ is completely fixed. All odd components of $N_{1234}$
vanish. The equations imply that $N_{1234}^{++++}$, and
$N_{1234}^{-+-+}$ are also zero (and so $N_{1234}^{+-+-}$, by mirror
symmetry). The non vanishing components can be written as
\begin{eqnarray}
N_{1234}^{--++} &=& \frac{1}{2}\left(I^1_{2,1,1,1}-I^1_{1,2,1,1}\right) ,
\nonumber \\
N_{1234}^{----} &=& 
\frac{1}{4}\left(I^0_{2,1,1,1}-I^0_{1,2,1,1}+I^0_{1,1,2,1}-I^0_{1,1,1,2}\right) ,
\label{eq:n:3.8}
\end{eqnarray}
together with
$N_{1234}^{--++}=N_{4123}^{+--+}=N_{3412}^{++--}=-N_{2341}^{-++-}$.

The expressions (\ref{eq:3.3a}), (\ref{eq:3.4}), (\ref{eq:n:3.7}) and
(\ref{eq:n:3.8}) have been obtained from $N_{12}^{\pm\pm}$,
$N_{123}^{\pm\pm\pm}$ and $N_{1234}^{\pm\pm\pm\pm}$ by fitting the
numerical coefficients in an expansion in terms of the
$I^\alpha_{r_1,\ldots,r_n}$. The fit is not unique and we have tried
to select the simplest ones.\footnote{In $N_{123}^{--+}$ we have
  allowed a factor $\sm_3^2$ in order to obtain a simpler expression.
  Of course, this is not mandatory since positive powers of $\sm_j^2$
  can be reabsorbed in the functions $I^\alpha_{r_1,\ldots,r_n}$.}  In
the next section we show that, by construction, the effective action
can always be written using the $I^\alpha_{r_1,\ldots,r_n}$ times
(possibly negative) integer powers of $\sm_j^2$. It is not clear why
these negative powers are actually not needed in the final expressions
for the effective action. It is noteworthy that this puzzle does not
exist for the current; when the method of covariant symbols is used
\cite{Salcedo:2000hx}, at no place $\sm^{-1}$ appears in the
calculation and by construction the final result involves only
functions $I^\alpha_{r_1,\ldots,r_n}$, as in (\ref{eq:3.3a}).

\subsection{Chern character -like ideas}
\label{subsec:3.C}

In this Section we take a small digression before going to the main
results of the present work.

It is tempting to try to find a simple systematics in the form of the
functions $N_{12}$, $N_{123}$ and $N_{1234}$ just quoted. However,
there is a number of problems to do that.  First, there is a huge
ambiguity in how the functions are expressed in terms of the
overcomplete basis $I^\alpha_{r_1,\ldots,r_n}$. Second, in $d=4$ there
is another ambiguity due to integration by parts (we have selected
$N_{123}^{-+-}=0$).  Finally, the expressions could perhaps display a
simple pattern if the redundant operators $\fF$ (in $d=2$) or $\fF^2$
(in $d=4$) were allowed in (\ref{eq:2.9}). In any case, we have been
unable to find any systematics in that expansion.  Nevertheless, there
is a remarkable exception: the expression
\begin{equation}
\Gamma=\langle \log(\sm^2+\sm \fm{}^\prime ) \rangle
 \,,
\label{eq:3.7}
\end{equation}
when expanded through order $d$, correctly reproduces the terms of the
type $(\sm \fm{}^\prime)^d$ of $W^-_c$, that is, the functions $N_{12}^{--}$
of $d=2$ and $N_{1234}^{----}$ of $d=4$. Indeed,
\begin{eqnarray}
\Gamma &=&
\left\langle 
\oint \frac{dz}{2\pi i} \log(z)\frac{1}{z-\sm^2-\sm \fm{}^\prime } 
\right\rangle
\nonumber \\
&=&
\left\langle 
I^0_1
+ I^0_{1,1} \sm \fm{}^\prime
+ I^0_{1,1,1} (\sm \fm{}^\prime)^2
+\cdots
\right\rangle
\nonumber \\
&=&
\left\langle 
I^0_1
+ \sm_1 I^0_{2,0} \fm{}^\prime
+ \sm_1 \sm_2 I^0_{2,1,0} \fm{}^\prime{}^2
+\cdots
\right\rangle
 \,.
\end{eqnarray}
Picking up the term which is a two-form, for $d=2$, and rewriting it
so that cyclic symmetry of the trace is manifest, produces
\begin{eqnarray}
\Gamma_{d=2} &=&
\left\langle 
\frac{1}{2} \sm_1 \sm_2 (I^0_{2,1}-I^0_{1,2}) \fm{}^\prime{}^2
\right\rangle
 \,,
\label{eq:3.9}
\end{eqnarray}
which is (\ref{eq:3.4}). In two dimensions this is the full result.
The corresponding result in $d=4$
\begin{eqnarray}
\Gamma_{d=4} &=&
\left\langle 
\frac{1}{4} \sm_1 \sm_2 \sm_3 \sm_4 
(I^0_{2,1,1,1}-I^0_{1,2,1,1}+I^0_{1,1,2,1}-I^0_{1,1,1,2}) \fm{}^\prime{}^4
\right\rangle
 \,,
\end{eqnarray}
is the full result when $v^R_\mu=v^L_\mu=0$ and $\fd\mrl=0$ (a case
studied in \cite{Salcedo:2000hx}), but not in general.  It is
noteworthy that (\ref{eq:3.7}) has some resemblance with the Chern
character of algebraic topology
\cite{Nakahara:1990th,Bertlmann:1996bk} which finds direct application
in anomalies and WZW actions \cite{Alvarez-Gaume:1985dr}.

In an attempt to reproduce all the LO components of $W_c^-$ we have considered
\begin{equation}
\Gamma=\langle \log(\sm^2+\sm \fm{}^\prime 
+{\cal O}_2+{\cal O}_3+{\cal O}_4+\cdots ) \rangle
 \,,
\end{equation}
where the ${\cal O}_n$ are general $n$-forms of the LO type:
\begin{eqnarray}
{\cal O}_2 &=& f^1_{123}\fm{}^\prime{}^2+f^2_{12}\fF \,,
\nonumber \\
{\cal O}_3 &=& f^3_{1234}\fm{}^\prime{}^3
+f^4_{123}\fm^\prime\fF +f^5_{123}\fF \fm^\prime\,,
\nonumber \\
{\cal O}_4 &=& f^6_{12345}\fm{}^\prime{}^4
+f^7_{1234}\fm^\prime{}^2\fF +f^8_{1234}\fm^\prime \fF \fm^\prime
+f^9_{1234}\fF \fm^\prime{}^2
+f^{10}_{123}\fF{}^2
\,,
\end{eqnarray}
etc. Here, the $f^k$ are functions of $\sm$ to be chosen suitably so
that $W^-_{c,{\rm LO}}=\Gamma$. Unfortunately, an analysis of the case
$d=4$ shows that no such functions exist if one requires them to be
(i) one-valued (no logarithms) and (ii) regular in the coincidence
limits $\sm_i^2\to\sm_j^2$. This statement holds regardless of how the
ambiguity in the functions $N_{123}$, $N_{1234}$ is fixed.

\section{Direct computation of the chiral remainder at LO}
\label{sec:4}


As discussed in Section \ref{subsec:n:2.C} the elegant method of Chan
\cite{Chan:1986jq} provides the derivative expansion of $\Tr\log K$
when the operator $K$ is of the form $M-D_\mu^2$, $M$ and $D_\mu$
being non abelian in general. Chan's technique is based on the symbols
method, which allows to write
\begin{equation}
\langle x|f(M,D_\mu)|x\rangle
=
\int \frac{d^d p}{(2\pi)^d} \langle x|f(M,D_\mu+p_\mu)|0\rangle
.
\label{eq:n:4.1}
\end{equation}
$f(M,D_\mu)$ denotes an operator constructed with $D_\mu$ (of the form
$\partial_\mu+A_\mu$) and $M$, a multiplicative operator, and we
assume $f$ to be sufficiently UV convergent.  On the other hand,
$|0\rangle$ is the state such that $\langle x|0\rangle=1$ (and so
$\partial_\mu|0\rangle=0$).\footnote{Thus, $\langle
  x|f(M,D_\mu+p_\mu)|0\rangle$ is just the standard symbol of the
  pseudo-differential operator $f(M,D_\mu)$.} Also by convenience, in
order to avoid the proliferation of $i$'s in the formulas, we use a
purely imaginary $p_\mu$, however, $p^2:=-p_\mu^2$ and $d^dp$ are the
usual ones.

The diagonal matrix element at the left-hand side of (\ref{eq:n:4.1})
is gauge covariant, but the matrix element at the right is not. The
expression becomes gauge covariant after taking the momentum
integration. This is easily understood as follows.  The momentum
integral is obviously invariant under the shift $p_\mu\to
p_\mu-a_\mu$, where $a_\mu$ is an arbitrary constant c-number. This
implies that, after momentum integration, the expression is invariant
under the shift $D_\mu\to D_\mu+a_\mu$. This implies that the operator
$D_\mu$ appears only through commutators, in the form $[D_\mu,~]$,
thus the expression is gauge covariant. (It is noteworthy that the
method of covariant symbols \cite{Pletnev:1998yu,Salcedo:2006pv}
provides gauge covariant expressions prior to momentum integration.)

In Chan's method, (\ref{eq:n:4.1}) is applied to $\log(M-D_\mu^2)$.
This gives \cite{Chan:1986jq}
\begin{eqnarray}
\Tr\log(M-D_\mu^2) &=& 
\int\frac{d^dx \, d^dp}{(2\pi)^d}
\tr\Big[
-\log \sN +
\frac{p^2}{d}\sN_\mu^2 
 \\
&&
+ 
\frac{4p^4}{d(d+2)}
\Big(
N_\mu^2N_\nu^2
 - \frac{1}{2}(N_\mu N_\nu)^2 
 - (N N_{\mu\mu})^2
 - 2 N F_{\mu\nu} NN_\mu N_\nu
 - \frac{1}{2} (F_{\mu\nu}N^2)^2
\Big)
+
\cdots
\Big] ,
\nonumber
\end{eqnarray}
where $ \sN=1/(p^2+M)$ and the dots refer to higher orders in the
derivative expansion. The method was extended to sixth order in
\cite{Caro:1993fs} and to curved space-time in \cite{Salcedo:2007bt}.

As shown in \cite{Salcedo:2008tc} and Section
\ref{subsec:n:2.C}, Chan's method combined with (\ref{eq:n:2.27})
allows to compute $W_c^-$ with explicit chiral covariance at every step.
However,  the use of Chan's
formula there implies a redefinition of the covariant derivative, by
the term $\sA_\mu$, cf. (\ref{eq:4.10}), which moreover involves two
Dirac matrices. Technically, this is quite inconvenient as the algebra
quickly produces rather long expressions. To sort this problem we
undertake here the task of developing an expansion from scratch along
Chan's ideas but specifically adapted to $W^-_c$.

To this end let us express (\ref{eq:n:2.27}) in the form
\begin{eqnarray}
W^-_c &=&
\frac{1}{2}\Tr\left[ \gamma_5 \log(\sm^2 -\thru\sD \,
\thru{\bar\sD}  )  \right],
\label{eq:4.16}
\end{eqnarray}
with
\begin{equation}
\bar\sD_\mu= \sm^{-1}\sD_\mu \sm=\sD_\mu+\sR_\mu 
,
\qquad
\sR_\mu = \sm^{-1}\sm_\mu
\,.
\end{equation}
An application of the method of symbols then gives
\begin{eqnarray}
W^-_c &=&
\frac{1}{2}\
\int\frac{d^dx \, d^dp}{(2\pi)^d}\, \tr\!\!\left[
 \gamma_5 \log(\sm^2 -(\thru\sD + \thrur{p}) \,
(\thru{\bar\sD}  + \thrur{p})  )  \right].
\label{eq:4.17}
\end{eqnarray}
Following \cite{Chan:1986jq}, the idea is to expand the logarithm in
terms ordered by the number of derivatives, using formal cyclic
symmetry of the trace, and then bring the expression to a manifestly
covariant form, that is, one where all $\sD_\mu$ operators (including
that in $\bar\sD_\mu$) appear only in commutators, $[\sD_\mu,~]$. Of
course, in order to guarantee that this works, UV divergences have to
be treated adequately. To this end the integrals over $p_\mu$ will be
dealt with using standard dimensional regularization. On the other
hand, because in the derivation of (\ref{eq:n:2.27}) it was assumed
that $\thru\sD$ anticommutes with $\gamma_5$ \cite{Salcedo:2008tc},
the Dirac gammas will be kept in the original integer dimension,
$d=2,4,\ldots$ (they are not dimensionally extended). As we show
subsequently this procedure allows to fix the UV ambiguity in such a
way that covariance is preserved.

Let us consider the second order contribution. Expanding the
logarithm and retaining terms with exactly
two derivatives gives
\begin{eqnarray}
W^-_{c,2} &=&
-\frac{1}{2}\
\int\frac{d^dx \, d^dp}{(2\pi)^d}\tr\!\left[
 \gamma_5 
\left(
\sN \!\thru\sD \,\thru{\bar\sD}
+\frac{1}{2}
\sN( \thru\sD \,\thrur{p}+ \thrur{p}\,\thru{\bar\sD} \,)
\sN( \thru\sD \,\thrur{p}+ \thrur{p}\,\thru{\bar\sD} \,)
\right)  \right]
\end{eqnarray}
where $\sN=1/(p^2+\sm^2)$. Rotational invariance of the integral over
$p_\mu$ then implies that an angular average can be taken, $p_\mu p_\nu\to
-p^2\delta_{\mu\nu}/d$,
\begin{eqnarray}
W^-_{c,2} &=&
-\frac{1}{2}\
\int\frac{d^dx \, d^dp}{(2\pi)^d}\tr\!\left[
 \gamma_5 
\left(
\sN \! \thru\sD \,\thru{\bar\sD}
-\frac{p^2}{2d}
\sN( \thru\sD \,\gamma_\mu + \gamma_\mu\,\thru{\bar\sD}\,)
\sN( \thru\sD \,\gamma_\mu + \gamma_\mu\,\thru{\bar\sD}\,)
\right)  \right].
\end{eqnarray}

At this point we can proceed to take the Dirac trace. Because the trace
with $\gamma_5$ requires at least $d$ gammas to give a non vanishing
result, and due to relations of type $\gamma_\lambda\gamma_\lambda=d$ and
$\gamma_\lambda\gamma_\alpha\gamma_\lambda=(2-d)\gamma_\alpha$, it is
clear that this second order contribution is identically zero when
$d>2$.  This is as expected for the second order term. In two
dimensions, using 
\begin{equation}
\gamma_\lambda\gamma_\alpha\gamma_\lambda=0 \,,\quad
\tr\gamma_5\gamma_\mu\gamma_\nu= -2i\epsilon_{\mu\nu} \qquad (d=2)
\end{equation}
gives
\begin{eqnarray}
W^-_{c,{\rm LO},d=2} &=&
\left\langle
\sN \fD \,\bar\fD
-\frac{2p^2}{d} \sN \fD \sN \bar\fD 
  \right\rangle_p,
\label{eq:4.22}
\end{eqnarray}
where 
\begin{equation}
\fD= \sD_\mu\,\fd x_\mu \,,\quad
\bar\fD= \bar\sD_\mu\,\fd x_\mu
\end{equation}
are 1-forms and $\langle~\rangle_p$ is short-hand for $i^{d/2}\!\int d^d
p/(2\pi)^d\,\tr(~)$ (including $x$-integration of the $d$-form).
Following \cite{Chan:1986jq}, we bring the expression to a more
homogenous form integrating by parts in momentum space. In the present
case
\begin{equation}
0= \
\int\frac{d^dp}{(2\pi)^d} \frac{\partial}{\partial p_\mu}( p_\mu \sN)
=
\int\frac{d^dp}{(2\pi)^d} (d\sN-2p^2\sN^2)
\end{equation}
allows to write
\begin{eqnarray}
W^-_{c,{\rm LO},d=2} &=&
 \left\langle
p^2
\left(
\sN^2 \fD \,\bar\fD - \sN \fD \sN \bar\fD 
  \right)
  \right\rangle_p
.
\label{eq:n:4.12}
\end{eqnarray}

Now, it can be verified that the expression is invariant under the
shift $\sD_\mu\to\sD_\mu+\sa_\mu$, $\bar\sD_\mu\to\bar\sD_\mu+\sa_\mu$
($\sa_\mu$ being a constant c-number). Therefore it is gauge
invariant. Indeed, it can be cast in a manifestly covariant form:
\begin{eqnarray}
W^-_{c,{\rm LO},d=2} &=&
\left\langle 
\frac{p^2}{2}
\left(
\sN^2 \{ \fD ,\bar\fD \} - [\fD, \sN][ \bar\fD ,\sN]
  \right)
\right\rangle_p.
\label{eq:4.26}
\end{eqnarray}

Let us note that $\int \tr\{ \fD ,\bar\fD \}$ vanishes,\footnote{We
  neglect topological contributions, such as $\int \tr \fF$ and assume
  the validity of formal integration by parts throughout.} hence the
momentum integral is actually convergent. Although $W^-_c$ is formally
logarithmically divergent, the requirement of covariance completely
fixes the UV ambiguities.\footnote{For this reason ambiguities in the
  procedure followed are not relevant in the final expression. For
  instance, taking the Dirac trace in $p_\mu
  p_\nu\gamma_\mu\gamma_\nu$ gives $-2p^2$, whereas taking the angular
  average first and then the trace gives $-4p^2/d$. The difference, of
  order $d-2$, vanishes in UV convergent expressions after $d\to 2$.}
This was also the case in the calculation of the covariant current
\cite{Salcedo:2000hx}.

Eq. (\ref{eq:4.26}) is the analogous of Chan's formula at second order
for the abnormal parity sector. The same expression can be written
using the more standard $\sF\!{}_{\mu\nu}$, $\sR_\mu$, $\sN$ and their
derivatives
\begin{eqnarray}
W^-_{c,{\rm LO},d=2} &=&
\left\langle  p^2 ( \sN^2  \fF + \fR  \sN  \fN^\prime \, )  \right\rangle_p .
\label{eq:4.27}
\end{eqnarray}
where
\begin{equation}
\fR= \sR_\mu\fd x_\mu = \sm^{-1}\fm^\prime
,
\qquad
\fN^\prime = N_\mu\fd x_\mu
.
\end{equation}

In order to bring (\ref{eq:4.27}) to the standard form in (\ref{eq:2.9})
we remove $\fF$ using the identity $\langle 2 p^2 N^2\fF\rangle_p =
\langle -N\fm^\prime N\fm^\prime\rangle_p $. Straightforward
manipulations yield then
\begin{eqnarray}
W^-_{c,{\rm LO},d=2} &=&
\left\langle 
\left(-\frac{1}{2}N_1 N_2  
+ p^2 (1-\sm_1^{-1}\sm_2 )N_1 N_2^2\right)\fm^\prime{}^2
 \right\rangle_p
\nonumber \\
&=&
\left\langle
\left(-\frac{1}{2}I^0_{1,1}  +  (1-\sm_1^{-1}\sm_2) I^1_{1,2}\right)\fm^\prime{}^2
 \right\rangle
.
\end{eqnarray}
Upon symmetrization using the cyclic property to enforce explicit mirror
symmetry, this result is easily shown to be equivalent to the known result
(\ref{eq:n:2.36}).

The relation analogous to (\ref{eq:n:4.12}) for the LO in $d=4$
dimensions is
\begin{eqnarray}
W^-_{c,{\rm LO},d=4}
&=&
\Big\langle
p^4
\Big(
-\frac{1}{3} \sN^2  \fD  \bar\fD  \sN^2  \fD  \bar\fD
-\frac{2}{3} \sN^3  \fD  \bar\fD  \sN  \fD  \bar\fD
-\frac{1}{3} \sN  \fD  \bar\fD  \sN  \bar\fD  \sN^2  \bar\fD
\nonumber \\ &&
-\frac{1}{3} \sN  \fD  \bar\fD  \sN^2  \bar\fD  \sN  \bar\fD
-\frac{1}{3} \sN  \fD  \sN  \fD  \sN^2  \fD  \bar\fD
-\frac{1}{3} \sN  \fD  \sN^2  \fD  \sN  \fD  \bar\fD
\nonumber \\ &&
+ \frac{2}{3} \sN  \fD  \sN^2  \bar\fD  \sN  \fD  \bar\fD
+ \frac{2}{3} \sN^2  \fD  \bar\fD  \sN  \fD  \sN  \bar\fD
-\frac{1}{3} \sN^2  \fD  \bar\fD  \sN  \bar\fD  \sN  \bar\fD
\nonumber \\ &&
-\frac{1}{3} \sN^2  \fD  \sN  \fD  \sN  \fD  \bar\fD
+ \frac{2}{3} \sN^2  \fD  \sN  \bar\fD  \sN  \fD  \bar\fD
-\sN  \fD  \sN  \fD  \sN  \fD  \sN  \fD
\nonumber \\ &&
+ \sN  \fD  \sN  \fD  \sN  \fD  \sN  \bar\fD
-\sN  \fD  \sN  \bar\fD  \sN  \fD  \sN  \bar\fD
+ \sN  \fD  \sN  \bar\fD  \sN  \bar\fD  \sN  \bar\fD
\Big)
 \Big\rangle_p .
\label{eq:4.29}
\end{eqnarray}
Once again this turns out to be covariant. Explicitly, the equation analogous
to (\ref{eq:4.27}) is
\begin{eqnarray}
W^-_{c,{\rm LO},d=4}
&=&
\Big\langle
p^4 \Big(
-\frac{2}{3} \fR  \sN^3  \fF  \fN^\prime
-\frac{1}{3} \sN^2  \fR^\prime  \sN^2  \fF
-\frac{1}{3} \sN^2  \fF  \sN^2  \fF
-\frac{2}{3} \sN^3  \fF  \sN  \fF
\nonumber \\ &&
-\frac{2}{3} \sN  \fR  \sN^2  \fN^\prime  \fF
-\frac{2}{3} \fR  \sN  \fR^\prime  \sN^2  \fN^\prime
-\frac{2}{3} \fR  \sN  \fF  \sN^2  \fN^\prime
+ \frac{1}{3} \fR  \sN^2  \fR  \fN^\prime  \fN^\prime
\nonumber \\ &&
-\frac{1}{3} \fR  \sN^2  \fR^\prime  \sN  \fN^\prime
-\frac{1}{3} \fR  \sN^2  \fF  \sN  \fN^\prime
+ \frac{1}{3} \sN^2  \fR  \fN^\prime  \sN  \fF
-\frac{1}{3} \sN  \fR  \sN  \fN^\prime  \sN  \fF
\nonumber \\ &&
+ \frac{1}{3} \sN  \fR  \sN  \fR  \fN^\prime  \fN^\prime
-\frac{1}{3} \sN  \fR  \sN  \fR^\prime  \sN  \fN^\prime
-\frac{1}{3} \sN  \fR  \sN  \fF  \sN  \fN^\prime
-\frac{1}{3} \sN  \fR  \sN^2  \fR  \sN  \fR^\prime
\nonumber \\ &&
+ \frac{1}{3} \fR  \sN  \fR  \sN  \fN^\prime  \fN^\prime
+ \frac{2}{3} \fR  \sN  \fR  \fN^\prime  \sN  \fN^\prime
+ \frac{1}{3} \fR  \sN^2  \fR  \sN  \fR  \fN^\prime
-\frac{2}{3} \sN^2  \fR  \sN  \fR  \sN  \fR^\prime
\nonumber \\ &&
+ \frac{2}{3} \sN  \fR  \sN  \fR  \sN  \fR  \fN^\prime
-\frac{1}{3} \fR  \sN  \fR  \sN  \fR  \sN  \fN^\prime
\Big)
 \Big\rangle_p .
\label{eq:4.30}
\end{eqnarray}

A technical comment is in order. Formal expressions of the type
(\ref{eq:n:4.12}) and (\ref{eq:4.29}) are essentially unique (the only
freedom being cyclic permutation). On the other hand, the gauge
invariant expressions (\ref{eq:4.27}) and (\ref{eq:4.30}) are by no
means unique, due to integration by parts and Bianchi identities. So
it is straightforward to go from the gauge invariant form to the
formal one by undoing all commutators but not the other way around.
To find a gauge invariant form as short as
possible  becomes a major issue \cite{Caro:1993fs,Salcedo:2004yh}.

To obtain (\ref{eq:4.30}) from (\ref{eq:4.29}) we have found it
convenient to write down the possible covariant terms and fit their
coefficients in order to reproduce (\ref{eq:4.29}), trying to minimize
the number of terms. (\ref{eq:4.30}) is not yet in the standard form
(\ref{eq:2.9}) but it can be brought to that form by eliminating terms
with $\fF^2$ by integration by parts. Then, labeling of operators
allows to directly use the basis functions
$I^\alpha_{r_1,\ldots,r_n}$.  We have verified that this procedure
precisely reproduces the results of Sec. \ref{sec:3}, obtained from
integration of the current.\footnote{To avoid errors, these
  manipulations and the similar ones at NLO in the next section, have
  been carried out with help of symbolic algebra software. The
  manipulations required are greatly simplified with the help of our
  notational conventions.} It should be noted that the right-hand side
of (\ref{eq:n:2.27}) is subject to UV ambiguities which could in
principle produce a spurious ``polynomial'' term of the form
$\langle\fF\fR^2\rangle$ \cite{Salcedo:2008tc}. Such term has not
appeared in the present calculation.

To finish this section, we note that these constructions, starting
from (\ref{eq:4.16}), show that $W^-_c$ can be
written (to all orders) using as building blocks $N$, $\sR_\mu$,
$\sF\!{}_{\mu\nu}$ and their covariant derivatives. Moreover, there is
a very definite pattern, characteristic of Chan's form, which is
illustrated by (\ref{eq:4.27}) and (\ref{eq:4.30}) at LO and by
(\ref{eq:5.7}) at NLO in next section, namely, at $n$-th order ($n$
derivatives) there are exactly $n$ blocks $N$. Compared to the bosonic
expansions in \cite{Chan:1986jq,Caro:1993fs}, in the fermionic case
the structure is complicated due to the presence of $\sR_\mu$, which
was absent in the bosonic case.  (Although some simplification is
gained due to the $d$-form structure as well as $\fF^\prime=0$ and
$\fX^{\prime\prime}=[\fF,\fX]$.) Because of the very restrictive pattern
allowed in expansions in Chan's form, the proliferation of terms is
avoided and this gives rise to compact expressions.

Another consequence of this constructive method, as compared with that
based on the current, is that it guarantees that $W^-_c$ can be
written using only terms of the form local operator (i.e., derivatives
of $\sm$ and $\sF\!{}_{\mu\nu}$) times $I^\alpha_{r_1,\ldots,r_n}$ times
integer powers of $\sm_j$ (the positive powers coming from derivatives
of $\sm^2$ and the negative powers coming from $\sm^{-1}$ in
$\sR_\mu$).

\section{NLO in two dimensions}
\label{sec:5}

\subsection{General considerations}
\label{subsec:5.A}

In this section we compute $W^-$ at NLO in $d=2$. Because the NLO is
UV finite, this coincides with $W^-_c$ at the same order. The general form
of this term is
\begin{eqnarray}
W^-_{{\rm NLO},d=2}=
 \left\langle
N^1_{12}\sm_{\alpha\alpha}\fF
+N^2_{123}\sm_\alpha^2\fF
+N^3_{123}\sm_{\alpha\alpha}\fm^\prime{}^2
+N^4_{1234}\sm_\alpha^2 \fm^\prime{}^2
\right\rangle ,
\label{eq:5.1}
\end{eqnarray}
where $N^1,N^2,N^3,N^4$ are four functions of the labeled $\sm$.
Mirror symmetry of $W^-$ implies the relations
\begin{equation}
N^1_{12}= -N^1_{21} 
\,,\quad
N^2_{123}= -N^2_{321} 
\,,\quad
N^3_{123}= -N^3_{21\m{3}} 
\,,\quad
N^4_{1234}= -N^4_{321\m{4}} 
\,.
\label{eq:5.2}
\end{equation}
In addition, $N^1$ and $N^3$ are odd functions of $\sm$, whereas $N^2$
and $N^4$ are even:
\begin{equation}
N^1_{12}= -N^1_{\m{1}\m{2}} 
\,,\quad
N^2_{123}= N^2_{\m{1}\m{2}\m{3}} 
\,,\quad
N^3_{123}= -N^3_{\m{1}\m{2}\m{3}} 
\,,\quad
N^4_{1234}= N^4_{\m{1}\m{2}\m{3}\m{4}} 
\,.
\label{eq:5.2b}
\end{equation}

All other possible structures not present in (\ref{eq:5.1}) are
redundant, either by using integration by parts or by rearrangement of
indices. The rearrangement comes from the identities
$X_{\alpha\beta}=X_{\beta\alpha}+[\sF\!{}_{\alpha\beta},X]$, and
\begin{equation}
\delta_{\alpha\beta}\epsilon_{\mu_1\mu_2\cdots\mu_d}=
\delta_{\alpha\mu_1}\epsilon_{\beta\mu_2\cdots\mu_d}
+\delta_{\alpha\mu_2}\epsilon_{\mu_1\beta\cdots\mu_d}
+\cdots
+\delta_{\alpha\mu_d}\epsilon_{\mu_1\mu_2\cdots\beta} \,.
\label{eq:5.3}
\end{equation}
This latter relation allows to transform any ``metric'' index into a
``differential form'' index. E.g.
\begin{equation}
\epsilon_{\mu\nu}\sF_{\alpha\mu} \sm_{\alpha\nu}
=
\epsilon_{\mu\nu}(\sF_{\mu\alpha} \sm_{\alpha\nu}
+\sF_{\nu\mu} \sm_{\alpha\alpha}
)
=
-\frac{1}{2}\epsilon_{\mu\nu}\sF_{\mu\nu} \sm_{\alpha\alpha}\,.
\end{equation}

Because integration by parts and rearrangement of indices preserve
the regularity condition, the functions $N^1,N^2,N^3,N^4$ are regular
in the coincidence limit. Unlike the LO of $d=4$, there are no
integration by parts ambiguities in (\ref{eq:5.1}). The only remaining
ambiguity comes from the two-dimensional identity
\begin{equation}
\sm_\alpha^2\fm^\prime{}^2=
\fm^\prime\sm_\alpha^2\fm^\prime -
\fm^\prime{}^2\sm_\alpha^2
- \sm_\alpha\fm^\prime{}^2 \sm_\alpha  \quad (d=2).
\end{equation}
This implies that the further condition
\begin{equation}
N^4_{1234}
-N^4_{\m{4}123}
+N^4_{\m{3}\m{4}12}
-N^4_{\m{2}\m{3}\m{4}1}
=0
\,,
\label{eq:5.5}
\end{equation}
can be imposed on $N^4$. (This choice is compatible with manifest
mirror symmetry (\ref{eq:5.2}).)  After $N^4$ is projected to comply
with this condition all functions are uniquely fixed.

To determine the functions $N^1,N^2,N^3,N^4$ one can use the method of
fixing them by reproducing the NLO current. The calculation of the
current can be done using the technique of covariant symbols, along
the lines explained in \cite{Salcedo:2000hx} for the LO.
Alternatively, the algebra involved in computing the covariant current
can be dealt with by using the world-line method
\cite{Strassler:1992zr,Schmidt:1993rk}.  This latter approach has been
applied in \cite{Hernandez:2007ng} to reproduce the LO results of
\cite{Salcedo:2000hx} and to the first computation of the NLO current
and of the effective action in $d=2$.  To carry out the same NLO
calculation we have applied two methods: first, one based on the
current, but computed using the method of covariant symbols.  This is
described in Section \ref{subsec:5.C}. And second, a direct computation of
the effective action using our version of Chan's approach in the
abnormal parity sector.  This approach is discussed subsequently. We
have verified that the two methods yield identical results and
moreover they are fully consistent with the results presented in
\cite{Hernandez:2007ng}.

\subsection{Direct computation of the effective action}
\label{subsec:5.B}

The Chan's -like derivative expansion technique of Sec.
\ref{sec:4} applies immediately to NLO in $d=2$.  In analogy with
(\ref{eq:4.27}) and (\ref{eq:4.30}), the manifestly covariant
expression in the present case is as follows:
\begin{eqnarray}
W^-_{{\rm NLO},d=2}
&=&
\Big\langle
p^4
\Big(
 \sN  \fR^\prime  \sN^2  \sN\!{}_{\alpha\alpha}
+ \sN^2  \sN\!{}_{\alpha\alpha}  \sN  \fF
+ \frac{1}{2} \sN^2  \fR^\prime  \sN^2  \sR_{\alpha\alpha}
+ \sN^2  \fF  \sN  \sN\!{}_{\alpha\alpha}
\nonumber \\ &&
+ \sN^3  \fF  \sN  \sR_{\alpha\alpha}
+ \frac{1}{2} \sN  \fN^\prime  \fN^\prime_\alpha  \sN  \sR_\alpha
+ \sN  \sN\!{}_\alpha  \sN\!{}_\alpha  \sN  \fF
-\sN  \fR  \fN^\prime  \sN  \sN\!{}_{\alpha\alpha}
\nonumber \\ &&
-\frac{1}{2} \sN  \fR  \fN^\prime  \sN\!{}_\alpha  \sN\!{}_\alpha
+ \frac{1}{2} \sN  \fR^\prime  \sN  \sN\!{}_\alpha  \sN\!{}_\alpha
-\sN  \fR^\prime  \sN^2  \sN\!{}_\alpha  \sR_\alpha
+ \frac{1}{2} \sN  \fR^\prime  \sN^2  \sR_\alpha  \sN\!{}_\alpha
\nonumber \\ &&
-\sN  \fF  \sN^2  \sN\!{}_\alpha  \sR_\alpha
-\frac{1}{2} \fN^\prime  \sN  \sR_\alpha  \sN  \fN^\prime_\alpha
+ \frac{1}{2} \fR  \sN  \sN\!{}_\alpha  \sN\!{}_\alpha  \fN^\prime
+ \fR  \sN  \sN\!{}_{\alpha\alpha}  \sN  \fN^\prime
\nonumber \\ &&
+ \fR  \sN  \sR_{\alpha\alpha}  \sN^2  \fN^\prime
-\frac{1}{2} \sN^2  \fR  \fN^\prime  \sN  \sR_{\alpha\alpha}
-\sN^2  \fF  \sN  \sN\!{}_\alpha  \sR_\alpha
+ \sN^2  \fF  \sN  \sR_\alpha  \sN\!{}_\alpha
\nonumber \\ &&
+ \sN  \sN\!{}_\alpha  \sN  \fF  \sN  \sR_\alpha
-\frac{1}{2} \sN  \fR  \sN  \fN^\prime  \sN  \sR_{\alpha\alpha}
+ \sN  \fR  \fN^\prime  \sN  \sN\!{}_\alpha  \sR_\alpha
-\frac{1}{2} \sN  \fR  \fN^\prime  \sN  \sR_\alpha  \sN\!{}_\alpha
\nonumber \\ &&
-\frac{1}{2} \sN  \fR  \fN^\prime  \sR_\alpha  \sN  \sN\!{}_\alpha
+ \sN  \fR  \sN\!{}_\alpha  \sN  \fN^\prime  \sR_\alpha
+ \frac{1}{2} \sN  \fR^\prime  \sN  \sR_\alpha  \sN  \sN\!{}_\alpha
-\frac{1}{2} \fR  \sN  \fN^\prime  \sN  \sR_\alpha  \sN\!{}_\alpha
\nonumber \\ &&
-\frac{1}{2} \fR  \sN  \sN\!{}_\alpha  \sR_\alpha  \sN  \fN^\prime
+ \fR  \sN  \fR  \sN\!{}_\alpha  \sN  \sN\!{}_\alpha
\Big)
 \Big\rangle_p .
\label{eq:5.7}
\end{eqnarray}
Here we have adopted the convention that derivatives with differential
form indices always act before derivatives with metric indices, that
is,
\begin{equation}
\fN^\prime_\alpha := (\fN^\prime)_\alpha=\sDa_\alpha\fN^\prime 
= \sN_{\alpha\mu} \fd x_\mu \,.
\end{equation}

We have selected the covariant terms in (\ref{eq:5.7}) in order to
obtain an expression as short as possible. In a sense, (\ref{eq:5.7})
is already the result, in compact form. Using integration by parts and
rearrangement of indices, the same functional can be brought to its
unique standard form (\ref{eq:5.1}).  A virtue of the present approach
(compared to that based on the current) is that it immediately
provides expressions for the functions $N^k$ that are manifestly
regular in the coincidence limit. Moreover they are linear
combinations of the type powers of $m_j$ times
$I^\alpha_{r_1,\ldots,r_n}$. 

In general, such expressions can be further simplified, as we do now.
Using the decomposition in components of well defined parity, as in
(\ref{eq:3.3}), we find (we use $N^1_{-+}$ for $N^{1,-+}_{12}$, etc):
\begin{eqnarray}
N^1_{-+} &=& -2  I^2_{3, 2}
 \,,
\nonumber \\
 N^2_{+++} &=& 2   I^3_{2, 1, 3} 
-2  I^3_{3, 1, 2} 
 \,,
\nonumber \\
N^2_{--+} &=& - I^2_{2, 2, 2} 
-4  I^2_{3, 1, 2} 
 \,,
\nonumber \\
N^2_{-+-} &=&
2  I^2_{2, 1, 3} 
-2  I^2_{3, 1, 2} 
 \,,
\nonumber \\
N^3_{-++}  &=&
2  I^2_{3, 2, 1} 
 \,,
\nonumber \\
N^3_{++-} &=&
   I^2_{2, 2, 2} 
 \,,
\nonumber \\
N^3_{---}  &=& 0
 \,,
\nonumber \\
N^4_{++++}  &=&
-2  I^3_{2, 1, 3, 1} 
+ 2 I^3_{3, 1, 2, 1} 
 \,,
\nonumber \\
N^4_{--++}  &=&
- I^2_{1, 3, 1, 2} 
-\frac{1}{2}   I^2_{2, 2, 1, 2} 
 + \frac{1}{2}   I^2_{2, 2, 2, 1} 
+ 3  I^2_{3, 1, 2, 1} 
 \,,
\nonumber \\
N^4_{-+-+}  &=&
  \frac{1}{2}   I^2_{1, 2, 2, 2} 
- I^2_{2, 1, 3, 1} 
-\frac{1}{2}   I^2_{2, 2, 1, 2} 
 +  I^2_{3, 1, 2, 1} 
 \,,
\nonumber \\
N^4_{-++-}  &=&
- I^2_{1, 2, 1, 3} 
 + \frac{1}{2}   I^2_{2, 1, 2, 2} 
-\frac{1}{2}   I^2_{2, 2, 1, 2} 
- I^2_{3, 1, 2, 1} 
 \,,
\nonumber \\
N^4_{+-+-} &=&
- I^2_{1, 2, 1, 3} 
 +  I^2_{1, 3, 1, 2} 
 + \frac{3}{2}   I^2_{2, 1, 2, 2} 
 + \frac{1}{2}   I^2_{2, 2, 2, 1} 
 \,,
\nonumber \\
N^4_{----}  &=& 0
 \,.
\end{eqnarray}
All other components not quoted follow from mirror symmetry
(\ref{eq:5.2}) or vanish by overall parity (\ref{eq:5.2b}).  Once
again we find that negative powers of $\sm_j^2$ are not required. We
do not have an explanation for this, but the simple expressions
obtained after simplification suggest that a more direct route could
exist.

The vanishing of $N^3_{---}$ and $N^4_{----}$ is easy to understand.
It follows from the observation in \cite{Salcedo:2000hx} (p.179) that,
for $d>0$, $W^-$ must vanish identically when one of the scalar
fields, e.g. $\mrl$, happens to be constant and there are no gauge
fields present:
\begin{equation}
W^-=0,\quad\hbox{whenever}\quad v^R_\mu=v^L_\mu= \fd\mrl=0 
\qquad (d>0)\,.
\end{equation}
At LO, this property dictates the form of $N_{12}^{--}$ in $d=2$ and
of $N_{1234}^{----}$ in $d=4$, in order to cancel the contribution of
$\Gamma_{\rm gWZW}$ \cite{Salcedo:2000hx}. At NLO, $W^-_c$ should
vanish by itself. As is easy to see, when $v^R_\mu=v^L_\mu=
\fd\mrl=0$, the only surviving contributions in (\ref{eq:5.1}) would
be those coming from $N^3_{---}$ and $N^4_{----}$ and so these
functions must vanish.

\subsection{Two dimensional NLO from the current }
\label{subsec:5.C}

For comparison we present here the calculation of the effective action
to NLO in two dimensions using the method based on the current.

The consistent current is defined as the variation of the effective
action under an infinitesimal change of the gauge field,
$\delta\sv_\mu$.  Specifically,
\begin{equation}
\delta W^- =
\langle \fJ_v^-\delta\fv \rangle
.
\end{equation}
Using the identity (\ref{eq:5.3}), the Lorentz index in
$\delta\sv_\mu$ can be taken as a differential form one so that
$\fJ^-_v$ is a $(d-1)$-form.

The covariant current at NLO takes the form
\begin{eqnarray}
\fJ^-_{v,{\rm NLO},d=2} &=&
A^1_{12}\fD_{\alpha\alpha}
+ A^2_{12} \fm^\prime_{\alpha\alpha}
+ A^3_{123} \sm_\alpha \fD_\alpha
+ A^3_{321} \fD_\alpha \sm_\alpha
\nonumber \\ &&
+ A^4_{123} \fm^\prime \sm_{\alpha\alpha}
- A^4_{321} \sm_{\alpha\alpha} \fm^\prime
+ A^5_{123} \sm_\alpha \fm^\prime_\alpha
- A^5_{321} \fm^\prime_\alpha \sm_\alpha
\nonumber \\ &&
+ A^6_{1234} \sm_\alpha \fm^\prime \sm_\alpha
+ A^7_{1234} \sm_\alpha^2 \fm^\prime
- A^7_{4321} \fm^\prime\sm_\alpha^2 \,.
\label{eq:A.1}
\end{eqnarray}
In this expression
\begin{equation}
\fD_\alpha= \sDa_\alpha\fD = \sF_{\alpha\mu} \fd x_\mu,
\quad
\fD_{\alpha\alpha} = \sF_{\alpha\alpha\mu} \fd x_\mu \,,
\end{equation}
and once again our convention is that derivatives with differential form
indices act before derivatives with metric indices, so
\begin{equation}
\fm^\prime_\alpha := \sDa_\alpha\fm^\prime\ = \sm_{\alpha\mu}\fd x_\mu \,, \quad
\fm^\prime_{\alpha\alpha} := \sm_{\alpha\alpha\mu}\fd x_\mu \,.
\end{equation}
In (\ref{eq:A.1}) we have used mirror symmetry to relate some of the
functions $A^k$.  In addition,
\begin{equation}
A^1_{12}=A^1_{21} ,\quad
A^2_{12}=-A^2_{21} ,\quad
A^6_{1234}=-A^6_{4321} .
\end{equation}
The explicit form of these functions is given below. They have been
obtained using the technique of covariant symbols
\cite{Pletnev:1998yu,Salcedo:2006pv} applied in \cite{Salcedo:2000hx}
to obtain the LO current. In \cite{Hernandez:2007ng} this NLO current
have been computed using the word-line approach. Note that the
functions in \cite{Hernandez:2007ng} are not identical to the $N$ and
$A$ here due to the different choice in the order of the derivatives.

Identifying the current (\ref{eq:A.1}) with the variation of
$W^-_{\rm NLO}$ in (\ref{eq:5.1}) yields the following set of
equations:
\begin{eqnarray}
A^1_{12} &=& (\sm_2-\sm_1  ) N^1_{12} 
- (\sm_1 + \sm_2) N^1_{2 \m{1}}
\,,
\nonumber \\
 A^2_{1 2} &=& -N^1_{1 2}
\,,
\nonumber \\ 
A^3_{1 2 3} &=& 
-2 N^1_{1 3} 
+ N^1_{3 \m{2}} 
+ (\sm_1 + \sm_3) (\nabla_2 N^1)_{\m{3} 1 2} 
+ (\sm_3-\sm_2) N^2_{1 2 3} 
- (\sm_1 + \sm_3) N^2_{\m{3} 1 2}
\,,\nonumber \\ 
A^4_{1  2  3} &=& 
-(\nabla_1 N^1)_{1  2  3} 
- (\sm_1 + \sm_3) N^3_{2  3  \m{1}} 
- (\sm_1 + \sm_3) N^3_{\m{3}  1  2}
\,,\nonumber \\ 
A^5_{1  2  3} &=&
-N^2_{1  2  3} 
+ (\sm_1 + \sm_3) N^3_{\m{3}  1  2}
\,,\nonumber \\ 
A^6_{1 2 3 4} &=& 
-(\nabla_2 N^2)_{1 2 3 4} 
- N^3_{\m{3}12} 
+ N^3_{\m{4}23} 
- (\sm_1 + \sm_4) (\nabla_1 N^3)_{\m{3}\m{4}12} 
- (\sm_1 + \sm_4) (\nabla_2 N^3)_{\m{4} 1 2 3} 
\nonumber \\ 
&&
+ (\sm_1 + \sm_4) N^4_{\m{3}\m{4}12}
+ (\sm_1 + \sm_4) N^4_{\m{4} 1 2 3}
\,,
\nonumber \\
A^{7}_{1  2  3  4} &=& 
-(\nabla_3 N^2)_{1  2  3  4} 
- N^3_{\m{4}  2  3} 
+ (\sm_1 + \sm_4) (\nabla_2 N^3)_{\m{4}  1 2 3} 
+ (\sm_1 + \sm_4) (\nabla_3 N^3)_{\m{4}123} 
\nonumber \\ 
&&
- (\sm_1 + \sm_4) N^4_{1  2  3  4} 
- (\sm_1 + \sm_4) N^4_{\m{4}123}
\,.
\label{eq:A.5}
\end{eqnarray}
In these equations $\nabla$ represents a variation operator that
increments the number of arguments by one
\cite{Salcedo:2000hx,Salcedo:2004yh}. Explicitly,
\begin{equation}
(\nabla_j f)(x_1,\ldots,x_n)= \frac{
f(x_1,\ldots,x_j,\widehat{x_{j+1}},\ldots,x_n)
-f(x_1,\ldots,\widehat{x_j},x_{j+1},\ldots,x_n)}{x_j-x_{j+1}} ,
\end{equation}
(where the hat indicates that the variable is missing). Note that
$\nabla_j$ represents the variation with respect the $j$-th argument
of $f$, so e.g.  $(\nabla_2 N^3)_{\m{4} 1 2 3}$ is not the variation
with respect to $\sm_2$, but the variation with respect the (abstract)
second argument of $N^3$, and the resulting function with four
arguments is then evaluated at $(-\sm_4,\sm_1,\sm_2,\sm_3)$.

The equations (\ref{eq:A.5}) have to be supplemented with mirror
symmetry of $W^-_{\rm NLO}$, (\ref{eq:5.2}), overall parity
(\ref{eq:5.2b}) and the condition (\ref{eq:5.5}). The full set of
equations is to be solved with respect to the unknowns $N^1$, $N^2$,
$N^3$, $N^4$ in terms of the known functions $A^k$.

$N^1$ is immediately obtained from $A^2$. The equation for $A^1$ is
automatically satisfied. $N^2$ and $N^3$ are obtained from $A^5$.
Indeed, exchanging the labels $1,3$ and using mirror symmetry produces
the new equation
\begin{equation}
A^5_{3  2  1} =
N^2_{1  2  3} 
+ (\sm_1 + \sm_3) N^3_{\m{3}  1  2}
\,.
\end{equation}
The two $A^5$ equations provide algebraic solutions for $N^2_{123}$
and $N^3_{\m{3}12}$. The equations $A^3$ and $A^4$ are
automatically satisfied.

Similarly, exchanging the labels $1,4$ and $2,3$ in $A^7$ and using
mirror symmetry gives a new $A^7$ equation. The two $A^7$ equations,
together with $A^6$ and the condition (\ref{eq:5.5}) provide algebraic
solutions to $N^4_{1234}, N^4_{\m{4}123}, N^4_{\m{3}\m{4}12},
N^4_{\m{2}\m{3}\m{4}1}$. One verifies that the four functions $N^4$ so obtained
are identical.

The solution found in this way coincides with that obtained in Sec.
\ref{subsec:5.B} through direct computation of the effective action.

For completeness we give below the functions $A^k$. The missing
components are related by mirror symmetry.
\begin{eqnarray}
A^1_{++} &=& 
-2 I^2_{2,2}
\,,
\nonumber \\
A^1_{--} &=& 0
\,,
\nonumber \\
A^2_{-+} &=&
2 I^2_{3,2}
\,,
\nonumber \\
A^3_{-++} &=&
-2 \sm_3^2 I^1_{2,1,2}
\,,
\nonumber \\
A^3_{+-+} &=&
-2 \sm_1^2 \sm_3^2 I^0_{2,1,2}
\nonumber \\
A^3_{++-} &=&
-2 I^1_{1,1,2}
\nonumber \\
A^3_{---} &=&
0
\,,
\nonumber \\
A^5_{+++} &=&
6 \sm_3^2 I_{1,1,4}^2+2 \sm_3^2 I_{1,2,3}^2
\,,
\nonumber \\
A^5_{--+} &=&
-2 \sm_3^2 I^1_{2,1,3}
\,,
\nonumber \\
A^5_{-+-} &=&
-2 \sm_3^2 I^1_{2,1,3}
\,,
\nonumber \\
A^5_{+--} &=&
-2 I_{1,2,3}^2-2 I_{1,3,2}^2-2 I_{2,1,3}^2-I_{2,2,2}^2
\,,
\nonumber \\
A^7_{+++} &=&
-4 \sm_3^2 I_{2,1,3}^2
\,,
\nonumber \\
A^7_{--+} &=&
-4 \sm_3^2 I^1_{2,1,3}
\,,
\nonumber \\
A^7_{-+-} &=&
-4 I_{2,1,3}^2
\,,
\nonumber \\
A^7_{+--} &=&
-4 I_{2,1,3}^2
\,,
\nonumber \\
A^9_{-+++} &=&
-2 I_{2,1,1,3}^3
-\frac{2}{3} I_{2,1,2,2}^3
+\frac{2}{3} I_{2,1,3,1}^3
+\frac{4}{3} I_{2,2,1,2}^3
+\frac{2}{3} I_{2,2,2,1}^3
+\frac{2}{3} I_{2,3,1,1}^3
+\frac{8}{3} I_{3,1,1,2}^3
\nonumber \\ &&
+\frac{4}{3} I_{3,1,2,1}^3
+\frac{4}{3} I_{3,2,1,1}^3+2 I_{4,1,1,1}^3
\,,
\nonumber \\
A^9_{+-++} &=&
-\frac{2}{3} I_{1,2,1,3}^3-\frac{2}{3} I_{1,2,2,2}^3-\frac{2}{3} \
I_{1,2,3,1}^3-\frac{4}{3} I_{1,3,1,2}^3
-\frac{4}{3} I_{1,3,2,1}^3-2 \
I_{1,4,1,1}^3-\frac{8}{3} I_{2,1,1,3}^3
\nonumber \\ &&
-\frac{4}{3} I_{2,1,2,2}^3-\frac{2}{3} \
I_{2,2,2,1}^3-\frac{4}{3} I_{2,3,1,1}^3+\frac{4}{3} I_{3,1,1,2}^3-\frac{2}{3} \
I_{3,2,1,1}^3
\,,
\nonumber \\
A^9_{+---} &=&
-4 I_{2,1,1,3}^2-2 I_{2,1,2,2}^2
\,,
\nonumber \\
A^9_{-+--} &=&
-4 I_{2,1,1,3}^2-2 I_{2,1,2,2}^2
\,,
\nonumber \\
A^{10}_{-+++} &=&
-\frac{2}{3} I_{2,1,1,3}^3+\frac{2}{3} I_{2,1,2,2}^3+2 \
I_{2,1,3,1}^3+\frac{4}{3} I_{2,2,1,2}^3+\frac{2}{3} I_{2,2,2,1}^3
-\frac{2}{3} I_{2,3,1,1}^3
\nonumber \\ &&
+\frac{8}{3} I_{3,1,1,2}^3
+\frac{4}{3} I_{3,1,2,1}^3-\frac{4}{3} \
I_{3,2,1,1}^3-2 I_{4,1,1,1}^3
\,,
\nonumber \\
A^{10}_{+-++} &=&
\frac{2}{3} I_{1,2,1,3}^3+\frac{2}{3} I_{1,2,2,2}^3+\frac{2}{3} \
I_{1,2,3,1}^3+\frac{4}{3} I_{1,3,1,2}^3+\frac{4}{3} I_{1,3,2,1}^3
+2 I_{1,4,1,1}^3+\frac{4}{3} I_{2,1,2,2}^3
\nonumber \\ &&
+\frac{8}{3} I_{2,1,3,1}^3
+\frac{8}{3} I_{2,2,1,2}^3
+2 I_{2,2,2,1}^3+\frac{4}{3} I_{2,3,1,1}^3+4 \
I_{3,1,1,2}^3+\frac{8}{3} I_{3,1,2,1}^3+\frac{2}{3} I_{3,2,1,1}^3
\,,
\nonumber \\
A^{10}_{++-+} &=&
-\frac{2}{3} I_{1,1,2,3}^3-\frac{4}{3} I_{1,1,3,2}^3-2 \
I_{1,1,4,1}^3-\frac{2}{3} I_{1,2,2,2}^3-\frac{4}{3} I_{1,2,3,1}^3-\frac{2}{3} \
I_{1,3,2,1}^3-\frac{4}{3} I_{2,1,1,3}^3
\nonumber \\ &&
+\frac{4}{3} I_{2,1,3,1}^3+\frac{4}{3} \
I_{2,2,1,2}^3+\frac{2}{3} I_{2,2,2,1}^3+\frac{8}{3} I_{3,1,1,2}^3+2 \
I_{3,1,2,1}^3
\,,
\nonumber
\\
A^{10}_{+++-} &=&
2 I_{1,1,1,4}^3+\frac{4}{3} I_{1,1,2,3}^3+\frac{2}{3} \
I_{1,1,3,2}^3+\frac{4}{3} I_{1,2,1,3}^3+\frac{2}{3} I_{1,2,2,2}^3+\frac{2}{3} \
I_{1,3,1,2}^3
\nonumber \\ &&
-\frac{8}{3} I_{2,1,1,3}^3-\frac{4}{3} I_{2,1,2,2}^3+\frac{2}{3} \
I_{2,2,1,2}^3+\frac{2}{3} I_{3,1,1,2}^3
\,,
\nonumber \\
A^{10}_{+---} &=&
-4 I_{2,1,1,3}^2-2 I_{2,1,2,2}^2
\,,
\nonumber \\
A^{10}_{-+--} &=&
-4 I_{2,1,1,3}^2-2 I_{2,1,2,2}^2
\,,
\nonumber \\
A^{10}_{--+-} &=&
-4 I_{2,1,1,3}^2-2 I_{2,1,2,2}^2
\,,
\nonumber \\
A^{10}_{---+} &=&
2 I_{2,1,2,2}^2+4 I_{2,1,3,1}^2+2 I_{2,2,1,2}^2+2 I_{2,2,2,1}^2+4 \
I_{3,1,1,2}^2+4 I_{3,1,2,1}^2
\,.
\end{eqnarray}

\section{Summary and conclusions}
\label{sec:6}

We have shown by direct calculation that, once the anomaly saturating
WZW term is subtracted from the effective action, the chiral invariant
remainder can be computed using a covariant formalism. Such a result
was available in the literature for the real part but not for the
imaginary part of the effective action.

The basic relation (\ref{eq:n:2.27}), holds to all orders, dimensions
and topologies and presumably can be extended to include gravitational
backgrounds. In particular, it should apply at finite temperature in
the imaginary time approach. As is known, there is a thermal chiral
invariant remainder \cite{Salcedo:1998tg} (the chiral anomaly is
temperature independent \cite{GomezNicola:1994vq}).

To carry out the calculations it has been extremely convenient to use
the notation introduced in \cite{Salcedo:2000hp,Salcedo:2000hx}. The
notation does not involve a rotation of the original fields appearing
in the Dirac operator. In this notation, chiral covariant expressions
behave formally as vector covariant ones and the number of structures
involved, and so the algebra, diminishes considerably.

In order to carry out a calculation within the derivative expansion,
we have found convenient to develop a new technique, along the lines
of Chan's approach for the bosonic case \cite{Chan:1986jq}. This
approach yields manageable expressions for the effective action.
Contributions to four derivatives are worked out explicitly and shown
to agree with the results obtained by integration of the current.

Finally, a suitable basis of functions is introduced in terms of which
the expressions, obtained by any method, are compactly packed while
being easily translatable to explicit form (explicit rational
functions with logarithms).  This basis appears naturally in our
calculation and avoids the need of fine-tuning required in the current
method to satisfy the regularity conditions noted in
\cite{Salcedo:2000hx}.

\acknowledgments
I thank C. Garc{\'\i}a-Recio for suggestions on the manuscript.
This work is supported in part by funds provided by the Spanish DGI
and FEDER funds with grant FIS2005-00810, Junta de Andaluc{\'\i}a
grants FQM225, FQM481 and P06-FQM-01735 and EU Integrated
Infrastructure Initiative Hadron Physics Project contract
RII3-CT-2004-506078.

\appendix

\section{Momentum integrals}
\label{app:A}

Let $I$ be the integral on the left-hand side of the relation
(\ref{eq:3.2}) . We assume $I$ to be UV convergent and $s+d/2-1$ to be
a non negative integer. In addition, the $r_j$ are integer and
$\sm_j^2>0$. After angular integration
\begin{eqnarray}
I
&=&
\frac{1}{(4\pi)^{d/2}\Gamma(d/2)}\int_0^\infty dx \,x^{s+d/2-1}
\prod_{j=1}^n\frac{1}{(x+\sm_j^2)^{r_j}}
.
\end{eqnarray}
This can be rewritten as
\begin{eqnarray}
I
&=&
\frac{(-1)^{s+d/2-1+\sum_j r_j}}{(4\pi)^{d/2}\Gamma(d/2)}\int_0^\infty dx 
\int_\gamma\frac{dz}{2\pi i} \left(\frac{1}{z_0+x}-\frac{1}{z+x}\right)
\nonumber \\
&& \times \, z^{s+d/2-1}\prod_{j=1}^n\frac{1}{(z-\sm_j^2)^{r_j}} \,.
\end{eqnarray}
Here $\gamma$ is a contour that starts at $-\infty$ (real) follows a
path just above the real negative axis reaching zero and then goes
back to $-\infty$ following a path just below the real negative axis.
For each $x$, this is equivalent to a closed negative contour
enclosing only the pole at $z=-x$. The term with $z_0$ has no pole and
so it gives no contribution.

Because the integral is UV convergent we can close the contour by
adding the contour at infinity. This closed path can then be deformed
to $\Gamma$, which encloses only the poles at $\sm_j^2$ (by assumption
there is no singularity at $z=0$).
\begin{eqnarray}
I
&=&
\frac{(-1)^{s+d/2-1+\sum_j r_j}}{(4\pi)^{d/2}\Gamma(d/2)}
\oint_\Gamma\frac{dz}{2\pi i} \log(z/z_0)
z^{s+d/2-1}\prod_{j=1}^n\frac{1}{(z-\sm_j^2)^{r_j}} \,.
\end{eqnarray}
In the UV convergent case this expression does not depend on the
subtraction point $z_0$. (In the term with $\log(z_0)$ $\Gamma$
encloses all the singularities and so it is equivalent to the contour
at infinity.) This is the right-hand side of (\ref{eq:3.2}).


\end{document}